\documentstyle[prd,aps]{revtex} \input epsf

\begin{document}
\title{
Excitation of the odd parity quasi-normal modes of compact objects}
\author{Zeferino Andrade and Richard H. Price}
\address{
Department of Physics, University of Utah, Salt Lake City, UT 84112}
\maketitle
\begin{abstract}
The gravitational radiation generated by a particle in a close
unbounded orbit around a neutron star is computed as a means to study
the importance of the $w$ modes of the neutron star.  For simplicity,
attention is restricted to odd parity (``axial'') modes which do not
couple to the neutron star's fluid modes. We find that for realistic
neutron star models, particles in unbounded orbits only weakly excite
the $w$ modes; we conjecture that this is also the case for
astrophysically interesting sources of neutron star perturbations.  We
also find that for cases in which there is significant excitation of
quadrupole $w$ modes, there is comparable excitation of higher
multipole modes.
\end{abstract}

 
\section{Introduction and overview}

Gravitational wave signals from perturbed black holes typically show
waveforms with strong quasinormal (QN) ringing, 
damped oscillations at a period and damping rate characteristic of the
mass and angular momentum of the hole.  If the source of perturbations
of the hole has no inherent time scale 
(as it would, for example, in the case of a particle  
scattered\cite{ON,O} or plunging into the hole\cite{DS}) then
QN ringing completely dominates the appearance of the waveform.
Indeed, most perturbation computations show waveforms that contain
only an initial transient and QN ringing. The meaning of QN ringing
has been well studied in the mathematical context of perturbation
calculations (Einstein's field equations linearized about a black hole
background). QN ringing, however, has also found to dominate the
waveforms of black hole processes computed with numerical solutions of
the full nonlinear set of Einstein's equations\cite{SP,GC}.

Unlike black holes, neutron stars have inherent time scales associated
with the various possible modes of fluid motion.  These fluid modes
can be understood, and analyzed approximately with a semi-Newtonian
approach: The excitations and frequencies of the fluid modes are
analyzed with Newtonian mechanics, and the fluid motions are treated
as a source in the equations of weak field general relativity.  The
emission of gravitational waves, in these calculations, damps the
periodic fluid oscillations of Newtonian theory, but due to the weak
coupling to gravitational waves these fluid QN modes are weakly
damped. By contrast, the QN modes of a Schwarzschild black hole are
strongly damped, with damping times of the same order as their period.

The sharp distinction between the nature of neutron star QN modes and
black hole QN modes ended when studies of neutron star models by
Kojima\cite{KOJI} and Kokkotas and Schutz\cite{KS1} revealed the
existence of modes that had no Newtonian counterpart.  These modes,
called $w$ modes (for wave modes) are general relativistic in their
origin and are similar to black hole QN modes in that they are quickly
damped. That $w$ modes are distinct from fluid modes is particularly
clear in the existence of odd-parity $w$ modes. The quantities
describing the perturbations of a spherical stellar model break up
into two sets of quantities, called even and odd parity, which are not
coupled by the perturbed field equations. The even parity set includes
all the information about fluid motion\cite{angmom}. The odd parity
set includes no fluid perturbations. It describes purely relativistic
modes that are perturbations only of the background geometry.

Chandrasekhar and Ferrari\cite{CHANDRAFER} showed that sufficiently
compact homogeneous neutron star models had odd parity QN modes. Later
studies \cite{AKK,AK2} not only confirmed these results but also
showed that these odd modes exist for all degrees of stellar
compactness and that the less compact the star is, the more rapidly
damped are its modes. Odd parity modes of neutron stars with 
realistic equations of state have also been computed recently
\cite{BBFER}.

In the next few years the vibrations of neutron stars may be of more
than academic interest. Neutron star processes, in various forms, are
among the most plausible sources of gravitational waves that may be
within the reach of detectors like LIGO, VIRGO, GEO600, TAMA300 and
others. Though the $w$ modes are at frequencies above the range of
high sensitivity for interferometric detectors\cite{KANDER}, they may
be interesting sources for resonant bar detectors.  In 1996, Andersson
and Kokkotas\cite{AK1} pointed out that much work on neutron star
dynamics was still being done in Newtonian theory and these studies,
in principle, excluded the fundamentally relativistic $w$
modes. Andersson and Kokkotas particularly directed attention to the
question that is most important about the $w$ modes: to what extent
are they excited in astrophysical processes. More specifically one
should ask about the relative excitation of the semi-Newtonian fluid
modes and the fundamentally relativistic $w$ modes in a real
astrophysical process.

The key
difficulty here lies in the question of what is an acceptable model of
a ``real astrophysical process.'' Those processes which are of greatest
interest as sources of detectable waves are the oscillations of a
neutron star formed in a supernova collapse or formed by the
coalescence of two neutron stars as the last stage of binary inspiral.
The numerical modelling of such processes with general relativistic
codes is beyond near term computational capability, so ``real'' processes
cannot be simulated. On the other hand it is not adequate to choose
any convenient source of perturbations. 

Andersson and Kokkotas (\cite{AK1}), for instance, gave an example of
the excitation of QN ringing of neutron stars by impinging
gravitational waves. It is plausible that the relative excitation of
$w$ modes and fluid modes for such a source is not representative of
how modes would be excited by astrophysical processes.  This might be
said also of any source of perturbations chosen arbitrarily. The model
we investigate here avoids arbitrariness by using a specific physical
process. We consider a nonrotating neutron star of mass $M$ and radius
$R$. An object of mass $m_{0}$ passes by the neutron star on a
geodesic trajectory that is the general relativistic equivalent of a
hyperbolic orbit. We make the ``particle'' approximation in which
$m_{0}\ll M$ and the field equations are expanded in the perturbation
parameter $m_{0}/M$. We study the dependence of excitation on the two
parameters of the particle orbits, the particle energy and angular
momentum. Neutron star models considered are limited to relativistic
polytropes. In principle, our ``incontrovertibly astrophysical'' model
would give a definitive answer to the question of the relative
excitation of fluid modes and $w$ modes. But in practice the present
paper gives only a partial answer and represents only a first
step. Here, for simplicity, we consider only odd parity modes. The
absence of coupling between odd parity perturbations and fluid motions
makes the odd parity case much simpler to analyze, but it also means
that there is no excitation of fluid modes with which comparisons can
be made. What can be achieved is a comparison of gravitational
radiation in $w$ modes, and gravitational radiation directly due to
the moving particle. In this sense the present work also shows the
parameter values for which $w$ mode excitation is significant. These
results are of interest by themselves and probably indicate the
systematics of the even parity excitations.

We present results for odd parity radiation from a range of particle
trajectories and we use these results to speculate about the
importance of $w$ modes in astrophysical processes.  We also
demonstrate that the study of the excitation of strongly damped QN
modes is better carried out with waveforms than with spectra.  An
additional point inherent in our results is that QN ringing of higher
multipoles is excited as strongly as that of quadrupole modes, though this
conclusion is probably a feature only of models involving
perturbations of a neutron star by a relativistic particle.

The remaining sections are organized as follows. Sec.\,\ref{model}
introduces the model problem. The wave equation describing the odd
parity perturbations appear in Sec.\,\ref{waveq}, the analysis of
the solution with Fourier transforms is given in Sec.\,\ref{FT}, and
the method of computing solutions is presented in
Sec.\,\ref{compimp}. Results are presented and discussed in
Sec.\,\ref{resdisc}, and conclusions are briefly summarized in
Sec.\,\ref{concl}. Some relevant details of particle orbits in the
Schwarzschild geometry are given in an appendix.  Throughout the paper
we use geometric units $G=c=1$ and metric signature $(+,-,-,-)$.

As this paper was in preparation we learned of two works dealing with
excitation of QN modes of a neutron star: a paper by Tominaga {\em et
al.}\cite{TOMSAIMAE} on substantially the same problem investigated
here, and a paper by Ferrari {\em et al}\cite{FERGUBO} which considers
the excitation of the even parity modes.  Among the differences
between these works and the present paper are an emphasis on waveforms
rather than spectra as a mean to detecting excitation of QN modes, the
study of high energy orbits and of odd modes of neutron stars of
radius greater than $3M$.

\section{The model: Perturbation of a static star by a particle moving in a scattered geodesic}\label{model}
In our model of gravitational interaction, we start with a static and
spherically symmetric spacetime background metric
\begin{equation}\label{metric}
ds^{2}=e^{\nu(r)}dt^{2}-e^{\lambda(r)}dr^{2}-r^{2}[d\theta^{2}+\sin^{2}\theta d\varphi^{2}]
\end{equation}
describing both the interior and exterior of a star of ideal fluid of mass M and radius $r=R$. 
The stress energy of the fluid is
\begin{equation}
T_{\alpha\beta}=(\rho+p)u_{\alpha}u_{\beta}-pg_{\alpha\beta}
\end{equation}
where $\rho$ is the mass-energy density and $p=p(\rho)$ is the pressure.
The mass function $m(r)$ is defined by $e^{-\lambda(r)}=1-2m(r)/r$,
and the structure of the stellar interior is found by solving the
hydrostatic equilibrium equations of general relativity
\begin{eqnarray}\label{hydrostatic}
& &\nonumber
\frac{dm}{dr}=4\pi r^{2} \rho\\
& &\nonumber
\frac{d\nu}{dr}=-\frac{2}{\rho+p}\frac{dp}{dr}\\
& &
\frac{dp}{dr}=-\frac{(\rho+p)[m(r)+4\pi r^{3} p]}{r[r-2m(r)]}\ .
\end{eqnarray}

We limit considerations here to stellar interiors that are
relativistic polytropes\cite{TOOPER}. The polytropic equation
of state has the simple algebraic form
\begin{equation}\label{polydef}
p/p_{0}=(\rho/\rho_{0})^{1+1/n}\ ,
\end{equation}
A particular model is specified by fixing the polytropic index, $n$, (in general
not an integer) and the central density and pressure $\rho_{0},p_{0}$.
With these fixed one can integrate to the surface (i.e, the radius at
which $p=0$) and find the value of the star's radius $R$, and mass $M=m(R)$.
The ratio $M/R$ is an indication of ``how relativistic'' the stellar
model is.
A particularly simple class of polytropics is the $n=0$ constant density
stars. For these models the metric coefficients are found to have
the closed forms
\begin{equation}\label{nuhomog}
e^{\nu(r)}=\left[\frac{3}{2}\sqrt{1-\frac{2M}{R}}-\frac{1}{2}\sqrt{1-\frac{2Mr^{2}}{R^{3}}}\right]^{2}\hspace*{20pt}  r\leq R
\end{equation}
\begin{equation}\label{lambdahomog}
e^{-\lambda(r)}=
\left(1-{2Mr^{2}}/{R^{3}}\right)\hspace*{20pt}  r\leq R\ .
\end{equation}
It is evident in (\ref{nuhomog}) that $e^{\nu(0)}$
vanishes for 
$M/R=4/9$. This well known limit to how relativistic 
a homogeneous model can be corresponds to infinite
central pressure. Outside the star, the metric functions 
take their familiar Schwarzschild form,
\begin{equation}
e^{-\lambda(r)}=e^{\nu(r)}=1-(2M/r)  \hspace*{10pt} r\geq R, 
\end{equation}

A point particle of mass $m_{0}\ll M$, starting from infinity, passes
close to the star (but never enters it), is scattered and proceeds out
to infinity. We treat the particle as a perturbation of the spacetime
inside and outside the star, and we analyze the perturbations only to
first order in $m_{0}/M$. The radiation reaction is second order in
this perturbation parameter, so we can ignore it and
assume that the particle  moves along a scattered geodesic of the unperturbed
spacetime exterior to the star, i.e. of Schwarzschild spacetime. The
metric of the perturbed spacetime can be written as
\begin{equation}
g_{\alpha\beta}=g^{(0)}_{\alpha\beta}+h_{\alpha\beta}\ ,
\end{equation}
where the ``(0)'' index  here and below denotes the background solution.
The Einstein equations to first order in perturbations are
\begin{equation}\label{Einsteins}
\delta R_{\alpha\beta}=8\pi[\delta T_{\alpha\beta}-\frac{1}{2}\delta T g^{(0)}_{\alpha\beta}-\frac{1}{2}T^{(0)}h_{\alpha\beta}]\ .
\end{equation}
where $T^{(0)}=\rho-3p$. The perturbed stress energy has two contributions. One contribution is
that of the fluid perturbations, a contribution we ignore for reasons
given below. The other contribution is the stress energy of the
scattered particle. We take the particle to be moving in the
($\theta=\pi/2$) equatorial plane on a timelike geodesic in which its
position as a function of coordinate time is given by
$z^\alpha(t)$. With this notation the particle stress energy in the coordinates 
(\ref{metric}) is given
by
\begin{equation}\label{stresspart}
\delta T^{\alpha\beta,particle}=m_{0}\frac{dt}{d\tau}\frac{\delta[r-R(t)]}{r^{2}}\delta[\cos\theta]\delta[\varphi-\varphi(t)]\frac{dz^{\alpha}}{dt}\frac{dz^{\beta}}{dt}\ .
\end{equation}
Here $R(t)$ and $\phi(t)$ are respectively the Schwarzschild radial
and azimuthal coordinates, as functions of Schwarzschild coordinate
time, for the particle orbit.

\section{The wave equation governing odd parity perturbations}\label{waveq}

A decomposition of  perturbation tensors in tensor spherical harmonics
(see \cite{ZERI}) leads to a split of 
(\ref{Einsteins}) into two sets of decoupled equations
\cite{THOCAMP} called even and odd parity.
Fluid perturbations are purely even parity and couple only to even
parity spacetime perturbations. Since we consider here only  the odd parity 
(sometimes also called axial) perturbations, our analysis 
involves only the stress energy perturbations of the particle itself, as
given in (\ref{stresspart})

We adopt the notation of Regge and Wheeler\cite{rw} and of
Moncrief\cite{moncrief}.  In this notation the only nonvanishing odd
parity metric perturbations, for a particular $l,m$ multipole, are
\begin{eqnarray}\label{pertdefs}
h_{t\theta}=-h_{0}^{lm}(r,t)(\partial/\sin{\theta}\partial\varphi)Y_{lm}\ \ \ \ h_{t\varphi}=h_{0}^{lm}(r,t)(\sin{\theta}\partial/\partial\theta)Y_{lm}\\
h_{r\theta}=-h_{1}^{lm}(r,t)(\partial/\sin{\theta}\partial\varphi)Y_{lm}\ \ \  
h_{r\varphi}=h_{1}^{lm}(r,t)(\sin{\theta}\partial/\partial\theta)Y_{lm}\\
h_{\theta\theta}=-h_{\varphi\varphi}/\sin^{2}{\theta}
=
h_{2}^{lm}(r,t)(\partial^{2}/\sin{\theta}\partial\theta\partial\varphi
-\cos{\theta}\partial/\sin^{2}{\theta}\partial\varphi
)Y_{lm}\\
h_{\theta\varphi}=
\textstyle{\frac{1}{2}}h_{2}^{lm}(r,t)(
\partial^{2}/\sin{\theta}\partial\varphi\partial\varphi
+\cos{\theta}\partial/\partial\theta
-\sin{\theta}\partial^{2}/\partial\theta\partial\theta
)Y_{lm}\ .
\end{eqnarray}
For simplicity, for now we make  the Regge-Wheeler\cite{rw} gauge choice, which 
for odd-parity means that we choose $h_{2}^{lm}=0$.
By taking a combination of the odd parity perturbed vacuum Einstein
equations, Regge and Wheeler\cite{rw} showed that $Q_{lm}(r,t)$
defined by
\begin{equation}\label{gauge}
Q_{lm}(r,t)\equiv e^{[\nu(r)-\lambda(r)]/2}\frac{h_{1}^{lm}(r,t)}{r}
\end{equation}
satisfies a simple homogeneous wave equation.  The same combination of
odd parity equations {\em with} stress energy perturbations\cite{ZERI}
results in a wave equation with a source term, of the form
\begin{equation}\label{ReggeWheeler}
\frac{\partial^{2}Q_{lm}}{\partial t^{2}}-\frac{\partial^{2}Q_{lm}}
{\partial r^{*2}}+\left\{l(l+1)-\frac{6m(r)}{r}+4\pi r^{2}[\rho(r)-p(r)]\right\}
\frac{Q_{lm}}{r^{2}}e^{\nu(r)}={\cal S}_{lm}(r,t)\ .
\end{equation}
Here $r^{*}$ is the usual {\it tortoise} coordinate
\begin{equation}
\frac{dr}{dr^{*}}=e^{[\nu(r)-\lambda(r)]/2}\ ,
\end{equation}
which, in the vacuum outside the star, can be written as
\begin{equation}
r^{*}=r+2M\log[r/2M-1]+{\rm const}\ .
\end{equation}
The source term, ${\cal S}_{lm}(r,t)$, is constructed from the stress
energy (\ref{stresspart}) of the particle. For a particle of mass $m_{0}$, 
moving in the
equatorial ($\theta=\pi/2$) plane, with energy per mass $E$, and
angular momentum per mass $L$, the $l,m$ multipole of the source term
has the form
\begin{equation}\label{source}
{\cal S}_{lm}(r,t)=\left(1-\frac{2M}{r}\right)\left(1-\frac{3M}{r}\right)\frac{2{\cal D}_{lm}}{r^{2}}-\frac{{\cal D}_{lm,r}}{r}\left(1-\frac{2M}{r}\right)^{2}+\frac{{\cal C}_{lm}}{r}\left(1-\frac{2M}{r}\right)
\end{equation}
where
\begin{equation}\label{defcalCnD}
{\cal D}_{lm} =D_{lm}\delta[r-R(t)] 
\hspace*{1in}{\cal C}_{lm}=C_{lm}\delta[r-R(t)]
\end{equation}
with
\begin{eqnarray}
& &\label{source1}
{D}_{lm}=\frac{16\pi i m m_{0}}{(l-1)l(l+1)(l+2)}\frac{L^{2}}{Er^{2}}\frac{\partial Y^{*}_{lm}}{\partial\theta}[\pi/2,\varphi(t)]\label{defD}\\
& &\label{source2}
{C}_{lm}=-\frac{16\pi m_{0}}{l(l+1)}\frac{dR}{dt}\frac{L}{r^{2}}
\frac{\partial Y^{*}_{lm}}{\partial \theta}[\pi/2,\varphi(t)]\label{defC}\ .
\end{eqnarray}
In the Schwarzschild spacetime outside the star
(\ref{ReggeWheeler}) 
is the well known Regge-Wheeler equation\cite{rw}.

\section{Fourier transform of the wave equation}\label{FT}

The equation in (\ref{ReggeWheeler}) could be solved directly with a
finite difference representation of the 1+1 partial differential
equation. Our Cauchy data for this equation, however, correspond to
the particle ``starting'' at infinity with no incoming radiation.
These starting conditions are most simply handled with a Fourier
representation of the problem, so we write $Q_{lm}(r,t)$ as a
Fourier integral
\begin{equation}
Q_{lm}(r,t)=\frac{1}{2\pi}\int_{-\infty}^{\infty}e^{-i\omega t}\widetilde{Q}_{lm}(r,\omega)
d\omega
\end{equation}
and transform (\ref{ReggeWheeler}) into a second order ODE for $\widetilde{Q}_{lm}
(r,\omega)$,
\begin{equation}\label{master}
\frac{d^{2}\widetilde{Q}_{lm}}{dr^{*2}}+\left\{\omega^{2}-
\left[\frac{l(l+1)}{r^{2}}-\frac{6m(r)}{r^{3}}+4\pi
(\rho(r)-p(r))\right]e^{\nu(r)}\right\}\widetilde{Q}_{lm}(r,\omega)=-\widetilde{S}_{lm}(r,\omega)
\end{equation}
where 
\begin{equation}
\widetilde{S}_{lm}(r,\omega)=\int_{-\infty}^{\infty}S_{lm}(r,t)e^{i\omega t} dt
\end{equation}
is the Fourier transform of the source term.

At infinity, 
the wave will be purely outgoing so that 
\begin{equation}\label{faraway}\begin{array}{cc}
\widetilde{Q}_{lm}(r,\omega)\rightarrow A_{lm}(\omega)e^{i\omega r^{*}}, & r^{*}\rightarrow \infty\ .
\end{array}
\end{equation}
It follows that at infinity, 
${Q}_{lm}$ is a function of retarded time $u\equiv t-r^{*}$ given by
\begin{equation}\label{waveatinfinity}
Q_{lm}(u\equiv t-r^{*})=\frac{1}{2\pi}\int_{-\infty}^{\infty}A_{lm}(\omega)
e^{-i\omega u}d\omega\ .
\end{equation}
It is useful to note that the definition of $Q_{lm}$ in terms of
multipole components of the real quantities $h_{ij}$, and the property
$Y_{lm}^{*}=(-1)^{m}Y_{l,-m}$ of the spherical harmonics, require
\begin{equation}
A_{lm}(\omega)=(-1)^{m}A^{*}_{l,-m}(-\omega)\ .
\end{equation}

The gravitational (odd parity part) power, radiated to infinity is given by
\cite{CUNPRIMO}
\begin{equation}\label{power}
P=\frac{1}{16\pi}\sum_{l=2}^{\infty}\sum_{m=-l}^{l}\frac{(l+2)!}{(l-2)!}
|Q_{lm}(u)|^{2}\ .
\end{equation}
Using (\ref{waveatinfinity}) and Parseval's theorem we obtain the odd parity
part of the energy spectrum
\begin{equation}\label{spectrum}
\frac{dE}{d\omega}=\frac{1}{16\pi^{2}}\sum_{l=2}^{\infty}\sum_{m=-l}^{l}\frac{(l+2)!}{(l-2)!}
|A_{lm}(\omega)|^{2}\ .
\end{equation}
From the wavefunction $Q_{lm}(u)$ we can construct the multipoles of the
metric perturbations. Of particular interest are the transverse
strains of gravitational waves. In an asymptotically flat gauge these
are the perturbations $h_{\theta\theta},h_{\phi\phi},h_{\theta\phi}$.
To compute the relationship between $Q_{lm}$ and perturbations in an
asymptotically flat gauge it is useful to note that the quantity
\begin{equation}\label{Qmonc}
Q^{{\rm Monc}
}_{lm}=\frac{1}{r}\left(
1-\frac{2M}{r}
\right)\left(
h_{1}^{lm}+\frac{1}{2}
\left[\frac{\partial h_{2}^{lm}}{\partial r}-\frac{2}{r}h_{2}^{lm}
\right]
\right)
\end{equation}
has been shown by Moncrief\cite{moncrief} to be a gauge invariant
combination of odd-parity metric perturbations in the Schwarzschild
background. In the stellar exterior, in the Regge-Wheeler gauge, the
perturbation $Q^{{\rm Monc}}_{lm}$ agrees with our wave function
$Q_{lm}$. It follows that $Q_{lm}$ is identical to $Q^{{\rm
Monc}}_{lm}$ in the stellar exterior, and that in the exterior 
we can take (\ref{Qmonc}) to be the definition of (\ref{gauge}) in an
arbitrary gauge.  Since the right hand side of (\ref{Qmonc}) is gauge
invariant, we can evaluate the relationship between $Q_{lm}$ and the
metric perturbations in an asymptotically flat gauge, at large
radius. In this case, the expression for $Q_{lm}$ in (\ref{Qmonc})
reduces to
\begin{equation}\label{asymptQ}
Q_{lm}=\frac{1}{2}\frac{\partial h_{2}^{lm}}{\partial r}+{\cal O}(r^{-2})
=-\frac{1}{2}
\dot{h}_{2,lm}
+{\cal O}(r^{-2})\ .
\end{equation}
Here the dot over ${h}_{2}^{lm}$ indicates the derivative with respect to 
retarded time $u$. The last equality in 
(\ref{asymptQ}) follows from the fact that ${h}^{lm}_{2}$ is a function 
of $u$ aside from corrections higher order in $1/r$.
From (\ref{asymptQ}) we have 
$\dot{h}_{2}^{lm}$ and hence, with (\ref{pertdefs}) we have, for example,
\begin{equation}
\dot{h}_{\theta\theta}=
\dot{h}_{2}^{lm}(r,t)(\partial^{2}/\sin{\theta}\partial\theta\partial\varphi
-\cos{\theta}\partial/\sin^{2}{\theta}\partial\varphi
)Y_{lm}\ .
\end{equation}
But this multipole component is complex. To get meaningful
physical quantities we need to sum multipole components with $\pm m$.
In doing this it is useful to define:
\begin{eqnarray}
F_{1}^{l,m}&=&\frac{1}{\sin\theta}\left[
\frac{\partial}{\partial\theta}-\cot\theta
\right]Y_{lm}(\theta,\varphi)\\
F_{2}^{l,m}&=&\frac{1}{2}\left[-\frac{m^{2}}{\sin^{2}\theta}
+\cos\theta\frac{\partial}{\partial\theta}-\sin\theta\frac{\partial^{2}}{\partial\theta^{2}}
\right]Y_{lm}(\theta,\varphi)\ .
\end{eqnarray}

For a given pair of multipole modes $l,m$ and $l,-m$ we can then construct
real functions 
\begin{eqnarray}\label{waves}
-\dot{h}_{\varphi\varphi}^{l,\Sigma m}/\sin^{2}{\theta}
=\dot{h}_{\theta\theta}^{l,\Sigma m}\equiv\dot{h}_{\theta\theta}^{l,m}+\dot{h}_{\theta\theta}^{l,-m}
&=&4|m|r\left[
{\rm Re}\left(Q^{lm}
\right)\sin{|m|\varphi}
+
{\rm Im}\left(Q^{lm}
\right)\cos{|m|\varphi}
\right]F_{1}^{l,|m|}\\
\dot{h}_{\theta\varphi}^{l,\Sigma m}\equiv
\dot{h}_{\theta\varphi}^{l,m}+\dot{h}_{\theta\varphi}^{l,-m}
&=&4r\left[
-{\rm Re}\left(Q^{lm}
\right)\cos{|m|\varphi}
+
{\rm Im}\left(Q^{lm}
\right)\sin{|m|\varphi}
\right]F_{2}^{l,|m|}\ .
\end{eqnarray}
\section{Computational implementation}\label{compimp}
\subsection{Solution for $A_{lm}(\omega)$}
If the background spacetime is due to a star, equation (\ref{master})
is to be solved for the boundary conditions of a regular solution at
the center of the star and outgoing waves at spatial infinity:
$Q_{lm}(r,\omega)\rightarrow r^{l+1}$ for $r\rightarrow 0$ and
$Q_{lm}(r,\omega)\rightarrow e^{i\omega r^{*}}$ for $r\rightarrow
\infty$. The Green function solution is found in the usual way. We
define $y^{\rm reg}_{l}(r,\omega)$ and $y^{\rm out}_{l}(r,\omega)$ as the
homogeneous solutions of (\ref{master}) with asymptotic forms
\begin{equation}\begin{array}{cc}
y^{\rm reg}_{l}(r,\omega)\rightarrow r^{l+1}, & r\rightarrow 0\\
y^{\rm out}_{l}(r,\omega)\rightarrow e^{i\omega r^{*}}, & r\rightarrow \infty\ .
\end{array}
\end{equation}
We then define the Wronskian of these two homogeneous solutions, an $r$
independent constant, to be
\begin{equation}\label{wronskstar}
W_{l}(\omega)=y^{\rm reg}_{l}\frac{dy^{\rm out}_{l}}{dr^{*}}-y^{\rm out}_{l}\frac{dy^{\rm reg}_{l}}{dr^{*}}\ .
\end{equation}
With the above definitions, the Green function solution is
\begin{eqnarray}
& &\nonumber
\widetilde{Q}_{lm}(r,\omega)=-\frac{1}{W_{l}(\omega)}\left[y^{\rm out}(r^{*},\omega)
\int_{r^{*}_{t}}^{r^{*}}\widetilde{S}_{lm}(r^{'*},\omega)y^{\rm reg}(r^{*'},\omega)dr^{*'}\right.\\
& &
\left.+y^{\rm reg}(r^{*},\omega)\int_{r^{*}}^{\infty}\widetilde{S}_{lm}(r^{*'},\omega)y^{\rm out}
(r^{*'},\omega)dr^{*'}\right]\label{Qint}\ ,
\end{eqnarray}
where the lower limit in the first integral, $r^{*}_{t}\equiv r^{*}(r_{t})$, 
denotes the 
value of $r^{*}$ at the minimum radius (turning point) of the scattered orbit.
In the limit of large $r$, a comparison of (\ref{faraway}) with (\ref{Qint}) 
gives us
\begin{equation}\label{amplitude}
A_{lm}(\omega)=-\frac{1}{W_{l}(\omega)}\int_{r^{*}(0)}^{\infty}
\widetilde{S}_{lm}(r^{*},\omega)y^{\rm reg}(r^{*},\omega)dr^{*}
\end{equation}

With the notation for ${\cal C}_{lm},{\cal D}_{lm}$ in (\ref{source}) -- (\ref{defcalCnD}) the expression for 
$\widetilde{S}_{lm}(r,\omega)$ takes the form
\begin{eqnarray}
& &\nonumber
\widetilde{S}_{lm}(r,\omega)=\int_{-\infty}^{\infty}dt e^{i\omega t}\left\{
\left[\left(1-\frac{2M}{r}\right)\left(1-\frac{3M}{r}\right)\frac{2D_{lm}}
{r^{2}}-\frac{D_{lm,r}}{r}\left(1-\frac{2M}{r}\right)^{2}\right.\right.\\
& &
\left.\left.+\frac{C_{lm}}{r}\left(1-\frac{2M}{r}\right)\right]\delta[r-R(t)]
-\frac{D_{lm}}{r}\left(1-\frac{2M}{r}\right)^{2}\frac{d\delta[r-R(t)]}{dr}
\right\}\ .
\end{eqnarray}
Substituting this expression into (\ref{amplitude}), and using the 
properties of the $\delta$ function, we get
\begin{eqnarray}
& &\nonumber
A_{lm}(\omega)=-\frac{1}{W_{l}(\omega)}\left\{
\int_{-\infty}^{\infty}dt e^{i\omega t}y^{\rm reg}[R,\omega]
\left[\left(1-\frac{2M}{R}\right)\frac{D_{lm}}{R^{2}}+\frac{C_{lm}}{R}\right]
\right.\\
& &
\left.+\int_{-\infty}^{\infty}dte^{i\omega t}\frac{dy^{\rm reg}}{dr}\frac{D_{lm}}{R}
\left(1-\frac{2M}{R}\right)\right\}\ .
\end{eqnarray}

For the geodesics considered in this work (scattered geodesics, see
appendix A), we can distinguish two phases of the motion: In the
first, the particle comes from infinity and moves towards the star
until reaching the turning point; it then returns to infinity in the
second region. If we choose the origin of coordinate time and angle
$\varphi$ at the turning point, then for any time $t=-\bar{t}$ and angle
$\varphi=-\bar{\varphi}$ in the first, inward, phase of the motion, there is
a corresponding point $t=\bar{t}, \varphi=\bar{\varphi}$
during the second, outward, phase.
The symmetry of the inward and outward motions allows us to write the 
amplitude as an integral only over positive time,
\begin{eqnarray}
& &\nonumber
A_{lm}(\omega)=-\frac{2i}{W_{l}(\omega)}\left\{\int_{0}^{\infty}y^{\rm reg}
(R,\omega)dt\left[\left(1-\frac{2M}{R}\right)\frac{|D_{lm}|}{R^{2}}
\cos(\omega t-m\varphi(t))+\frac{|C_{lm}|}{R}\sin(\omega t-m\varphi(t))\right]
\right.\\
& &\label{amplitude2}
\left.+\int_{0}^{\infty}dt\frac{dy^{\rm reg}}{dt}\frac{|D_{lm}|}{r}
\left(1-\frac{2M}{R}\right)\cos(\omega t-m\varphi(t))\right\}\ ,
\end{eqnarray}
where $|D_{lm}|=-iD_{lm}{exp(im\varphi)},|C_{lm}|=C_{lm}exp(im\varphi)$ with
$D_{lm},C_{lm}$ given by (\ref{source1}-\ref{source2}).
Once the quantities $A_{lm}(\omega)$ are computed, the waveform
follows from (\ref{waves}) and the energy spectrum from
(\ref{spectrum}).

If the particle moves outside a spherically symmetric black hole, not
a star, then the amplitude can be obtained from (\ref{amplitude2}) with
only two changes.  The ingoing solution $y^{ing}$ (see
\cite{LOUPRICE}), of the Regge-Wheeler equation must be substituted
for $y^{\rm reg}$ in the integral in (\ref{amplitude2}), and the
Wronskian defined in (\ref{wronskstar}) must be replaced by
\begin{equation}\label{wronskhole}
W_{l}(\omega)=y^{\rm ing}_{l}\frac{dy^{\rm out}_{l}}{dr^{*}}-y^{\rm out}_{l}\frac{dy^{\rm ing}_{l}}{dr^{*}}\ .
\end{equation}

\subsection{Numerical method}\label{nummethod}
The first step in the determination of the amplitude and waveform is the computation of the regular homogeneous solution
of the equation (\ref{master}). The numerical problem can be divided in two parts: Integration of (\ref{master}) from the center of the star to its surface. And then from the surface to the point (outside the star) where we want to compute $y^{\rm reg}(r)$. Due to the presence of the centrifugal potential, which diverges at the center of the star, it is convenient to introduce, inside the star, the new function $Z_{l}$, through
\begin{equation}
y^{\rm reg}_{l}=r^{l+1}Z_{l}(r)
\end{equation}
and instead of (\ref{master}), integrate 
\begin{eqnarray}
& &\nonumber
\frac{d^{2}Z_{l}}{dr^{2}}+\left\{\frac{2(l+1)}{r}+e^{\lambda(r)}\left[\frac{2m(r)}{r^{2}}+4\pi r(p-\rho)\right]\right\}\frac{dZ_{l}}{dr}+\\
& &\label{inside}
\left[\omega^{2}e^{-\nu(r)}+\frac{(8-2l^{2})m(r)}{r^{3}}+4\pi[p(r)-\rho(r)](l+2)\right]e^{\lambda(r)}Z_{l}=0
\end{eqnarray}
from a point $r_{0}$, near the center of the star, using the starting conditions
\begin{eqnarray}
& &
Z_{l}(r_{0})=1+\frac{r^{2}_{0}}{2l+3}\left\{[(2l+3)l-2]\frac{2\pi}{3}\rho_{0}-\frac{\omega^{2}e^{-\nu_{0}}}{2}-2\pi(l+2)p_{0}\right\}\\
& &
\frac{dZ_{l}}{dr}(r_{0})=\frac{2r_{0}}{2l+3}\left\{[(2l+3)l-2]\frac{2\pi}{3}\rho_{0}-\frac{\omega^{2}e^{-\nu_{0}}}{2}-2\pi(l+2)p_{0}\right\},
\end{eqnarray}
until reaching the surface of the star, $r=R$ where we switch back 
to $y^{\rm reg}_{l}$ and integrate
\begin{equation}\label{outside}
\frac{d^{2}y^{\rm reg}_{l}}{dr^{2}}+\frac{1}{1-2M/r}\frac{2M}{r^{2}}\frac{dy^{\rm reg}_{l}}{dr}+
\frac{1}{(1-2M/r)^{2}}\left\{\omega^{2}-\left(1-\frac{2M}{r}\right)\left[\frac{l(l+1)}{r^{2}}-\frac{6M}{r^{3}}\right]\right\}y^{\rm reg}_{l}=0
\end{equation}
from $r=R$, with the starting conditions,
\begin{eqnarray}
& &
y^{\rm reg}_{l}(R)=R^{l+1}Z_{l}(R)\\
& &
\frac{dy^{\rm reg}_{l}}{dr}=R^{l+1}\frac{dZ_{l}}{dr}(R)+(l+1)R^{l}Z_{l}(R)
\end{eqnarray}
until the desired point $r$.

Both (\ref{inside}) and (\ref{outside}) follow from a rewriting of
(\ref{master}) in terms of the variable $r$. This avoids the
numerically time consuming task of finding $r(r^{*})$.

If the unperturbed spacetime is a Schwarzschild black hole instead of
a star, then instead of $y^{\rm reg}_{l}$ we need to find $y^{\rm ing}_{l}$,
the purely ingoing (at the horizon) homogeneous solution of
(\ref{master}). This is done by substituting $y^{\rm reg}_{l}$ by $y^{\rm
ing}_{l}$ in (\ref{outside}) and integrating from a large negative value
of $r^{*}$ with the ``initial'' condition $y^{\rm ing}_{l}\sim e^{-i\omega
r^{*}}$.

The computation of the Wronskian in
(\ref{wronskstar}) or (\ref{wronskhole}) is now straightforward.  We simply
compute $y^{\rm reg}_{l}$ (or $y^{\rm ing}_{l}$) at large values of $r$ and
use the asymptotic form of $y^{\rm out}_{l}$ at large r
\begin{equation}
y^{\rm out}_{l}(r\rightarrow\infty)\sim e^{i\omega r^{*}}\left[1-\frac{l(l+1)}{2i\omega r}\right]
\end{equation}
to find the Wronskian by definition.

One point worth of mention is that since
\begin{equation}
\frac{\partial Y^{*}_{lm}}{\partial\theta}[\pi/2,0]=-\sqrt{\frac{2l+1}{4\pi}\frac{(l-m)!}{(l+m)!}}\frac{2^{m+1}}{\sqrt{\pi}}\frac{\Gamma[1+(l+m)/2]}{\Gamma[1/2+(l-m)/2]}\sin[\frac{\pi}{2}(l+m)],
\end{equation}
where $\Gamma$ denotes the Gamma function, $A_{lm}\neq 0$, for a given $l$,
only when $m=-l+1,-l+3,...,l-1$. 
Thus for $l=2$ only $A_{21},A_{2-1}$ are non zero and for $l=3$ only $A_{30},A_{32},A_{3-2}$ are non zero. 

A consequence is that if, as we will do in the next section, we restrict 
attention to the waveform seen by an observer in the plane where the
particle moves, then $\dot{h}^{l,\sum m}_{\theta \varphi}=0$ in (\ref{waves})
and the waveform will be polarized in the direction
$\varphi,\varphi$.

\section{Numerical Results and Discussion}\label{resdisc}


Our presentation of results starts with gravitational radiation in the
quadrupole ($l=2$) and octupole ($l=3$) modes, for a particle moving
around a nonrotating black hole (see also \cite{ON},\cite{O}). Though
our main interest here is neutron star $w$ modes, it will be useful to
establish the black hole results for comparison. The case of a black
hole will also be helpful in clarifying several issues relevant to the
neutron star case.  An additional motivation for the presentation of
these results is to demonstrate that in cases in which $l=2$ QN
ringing is important, the $l=3$ ringing tends to be comparable, both
for black holes and neutron stars.


It will be useful to understand the aspects of the emerging radiation
that are associated with the motion of the particle, and those that
may be ascribed to the influence of the background spacetime. (There
is, of course, no fully satisfactory way of making such a
distinction.) 
If the particle were in a periodic orbit with frequency $\Omega$, its
gravitational perturbations would be functions of $\varphi-\Omega t$,
and $l,m$ multipoles would have time dependence only at frequency
$m\Omega$.  For the scattering orbits one would expect that the
characteristic peak of the spectrum would follow an approximate form
of this rule with the particle's angular velocity $d\varphi/dt$ at the
turning point $r_{t}$ taking the place of $\Omega$. In the discussion
below we will use the expression ``particle frequency'' to denote this
nonrelativistically expected peak frequency $m\,d\varphi/dt$.  Note
that, as discussed above, an orbit in the equatorial plane is a source
only for multipoles with $m=l-1, l-3,\cdots,-l+1$. Thus, $m$ will be 1 for
quadrupole radiation and 2 for octupole.

\begin{figure}
\hspace*{.1\textwidth}
\epsfxsize=0.35\textwidth
\epsfbox{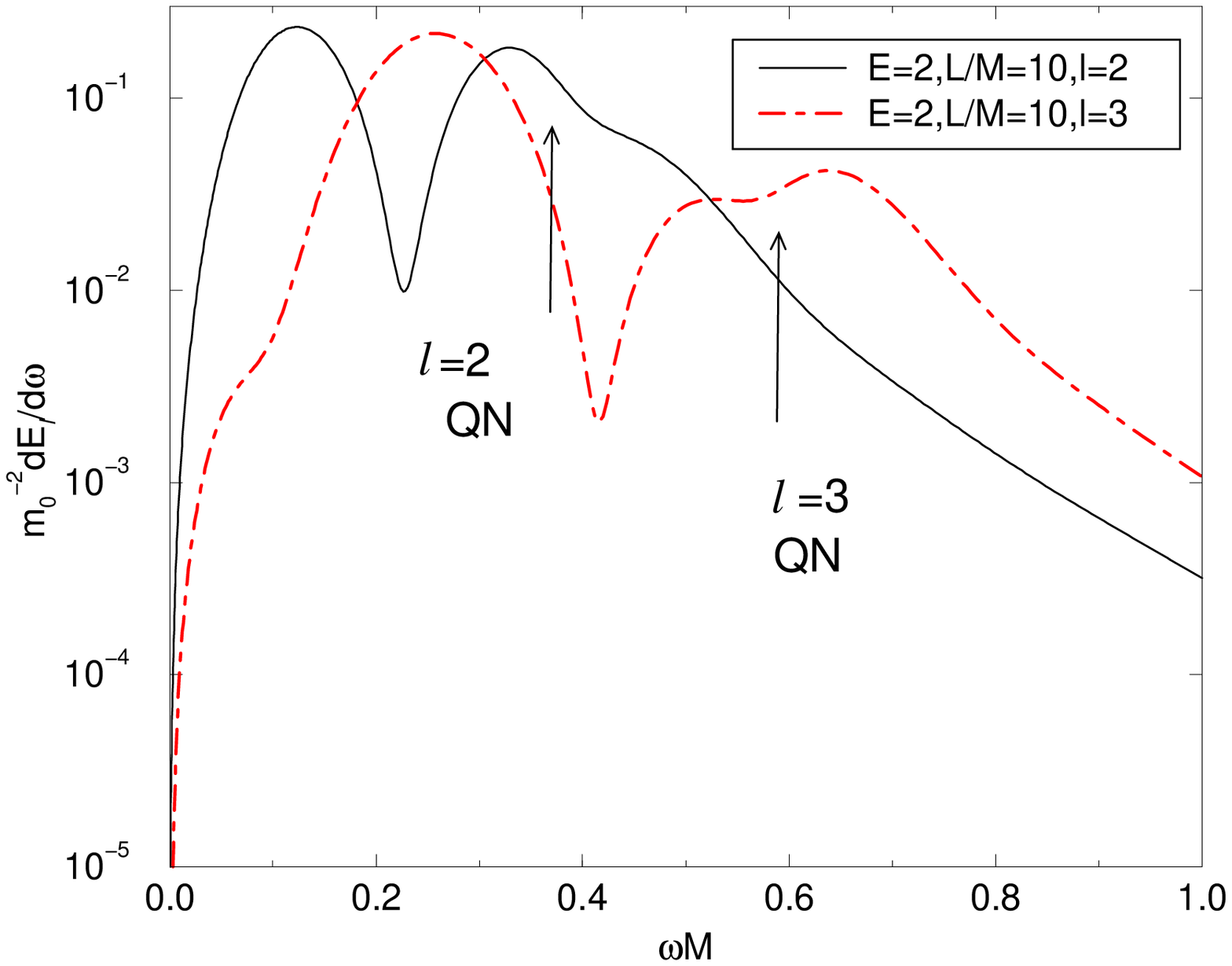}\hspace*{.1\textwidth}\epsfxsize=0.35\textwidth
\epsfbox{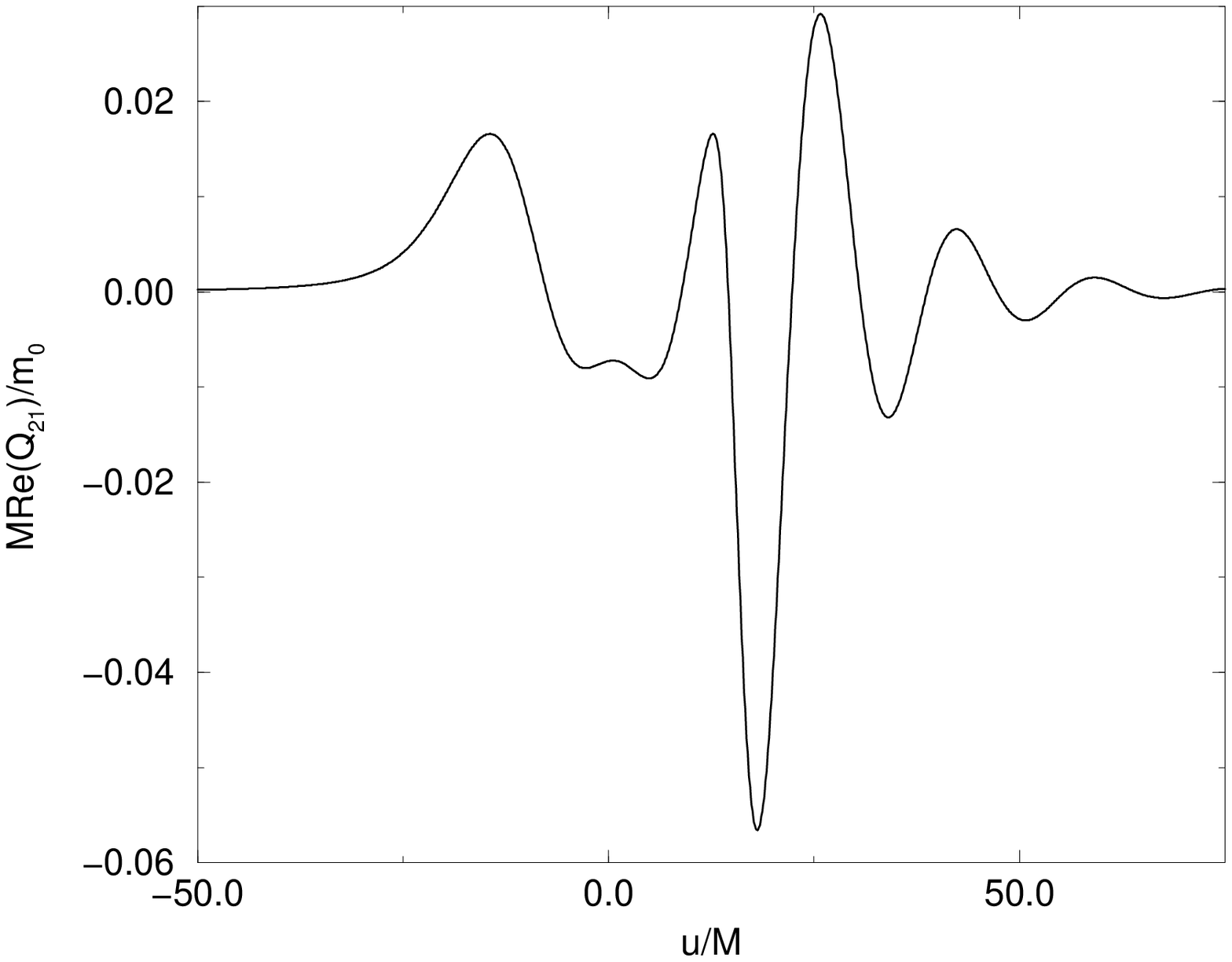}
\caption{\label{BHspect} Gravitational
radiation from a particle scattered by a nonrotating black
hole.  The particle's energy per mass is $E=2$ and its angular
momentum per mass is $L=10$, so that the orbital turning point
(minimum radius) is $r_{t}=10M/3$.
(a) The energy spectrum for $l=2$ and $l=3$ radiation.
Arrows indicate the oscillation frequency (i.e.\ , real part of the QN
frequency) for the least damped $l=2$ mode ($\omega M=.374$)
and the least damped
$l=3$ ($\omega M=.599$) modes. (b) The $l=2$ waveform in the equatorial ($\theta=\pi/2$) 
plane. The real part of the wavefunction $Q_{21}$ is shown; this corresponds
to the waveform seen at azimuthal angle $\varphi=\pi/2$. 
}
\end{figure}

Figures \ref{BHspect}\,a and b show $l=2$ and $l=3$ spectra and
waveforms for orbits with particle energy parameter $E$ (energy per
particle mass) of 2.  For the particle motion in Fig.~\ref{BHspect}
the angular frequency $d\varphi/dt$ of the particle at the turning
point is $0.18/M$ so that the particle frequency is $0.18/M$ for
quadrupole radiation and $0.36/M$ for octupole.   The waveform
in Fig.~\ref{BHspect}b shows more specifically that QN ringing is an
important part of the waveform.

 The locations of the real parts of the least damped QN modes, $\omega
M=.374$ for $l=2$ and $\omega M=.599$ for $l=3$, are indicated with
arrows in Fig.~\ref{BHspect}\,a.  The analogy with normal mode systems
would suggest that the extent to which particle motion stimulates QN
ringing increases the closer the characteristic particle frequency is
to the QN frequency.  In the case of Fig.~\ref{BHspect}\,a the (odd
parity $m=2$) particle frequency for $l=3$ case is twice that for the
(odd parity $m=1$) $l=2$ case, and the $l=3$ QN frequency is around
60\% higher than the $l=2$ QN frequency. Both particle frequencies are
roughly equal in the suitability for exciting the respective QN modes,
and the plots show that QN ringing seems to be of roughly equal
importance for $l=2$ and $l=3$.

\begin{figure}
\hspace*{.1\textwidth}\epsfxsize=0.8\textwidth
\epsfbox{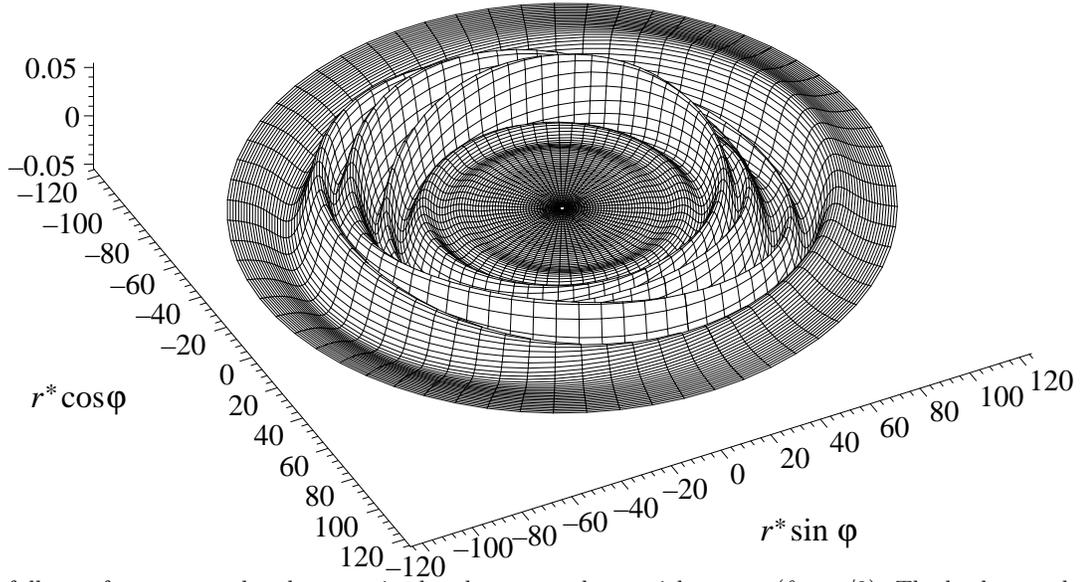}
\caption{\label{HomogstarR=3.1253Dwform} The full waveform
as seen by observers in the plane were the particle moves
($\theta=\pi/2$). The background spacetime is a homogeneous star of radius
$R=3.125M$ and the energy of the particle is $E=2$ with an angular
momentum $L=10M$. The waveform as seen in the same plane in the
direction $\varphi=\pi/2$ is shown below in Fig.~5(b). The middle part
(spiral part) represents the ringing part of the wave. The inner
(center) the tail. The outer part is the initial burst. In the figure,
the $\varphi=0$ symmetry axis of the particle orbit is the line $y=r^{*}
\sin\varphi=0$. }
\end{figure}

The presence of QN radiation is fairly clear in the waveform, shown in
 Fig.~\ref{BHspect}\,b. This figure shows the gravitational radiation,
 a mixture of effects of particle radiation and QN ringing. It does
 not tell us what fraction of the radiated energy is in QN ringing,
 but that question probably cannot be given a clear formulation in any
 case\cite{nollertprice}.  As an indicator of the importance of QN
 ringing, a plot of the waveform has a clear advantage, but it also
 has a disadvantage. To show the waveform \ref{waves} we must choose a
 direction in the $\theta,\varphi$ 2-sphere. The curve in
 Fig.~\ref{BHspect}\,b corresponds to the equatorial plane
 ($\theta=\pi/2 $) and to $\varphi=\pi/2$. The appearance of the
 waveform in other directions will differ.  As an indication of the
 complexity of the waveform dependence on direction, a quadrupole
 waveform in the equatorial plane is shown in
 Fig.~\ref{HomogstarR=3.1253Dwform}. The radiation is that for a
 particle with $E=2$ and $L=10M$ moving in the field of a constant
 density star with $R=3.125M$.  The waveform amplitude, ${\rm
 Re}(Q_{21})\sin\varphi +$ ${\rm Im}(Q_{21})\cos\varphi$, is plotted
 above a cartesian plane with $x\equiv r^{*}\cos\varphi$, $y\equiv
 r^{*}\sin\varphi$.  The direction of the symmetry axis of the
 particle orbit is the straight line $y=0$ in the figure. Some
 symmetries of the radiation pattern can be seen in
 Fig.~\ref{HomogstarR=3.1253Dwform}. Since the radiation is $m=1$
 radiation, the amplitude at a given time (i.e.\ , at a given value of
 $r^{*}$) must be antisymmetric under
 $\varphi\rightarrow\varphi+\pi$. At each radius in
 Fig.~\ref{HomogstarR=3.1253Dwform}, then, the variation in amplitude
 must have two nodes. It is interesting to note that the particle
 trajectory is a symmetric one (it has a line of bilateral symmetry)
 but the radiation pattern is not symmetric since there is a delay in
 time between the radiation given off by the initial and final
 portions of the particle trajectory.

\begin{figure}
\hspace*{.1\textwidth}\epsfxsize=0.35\textwidth
\epsfbox{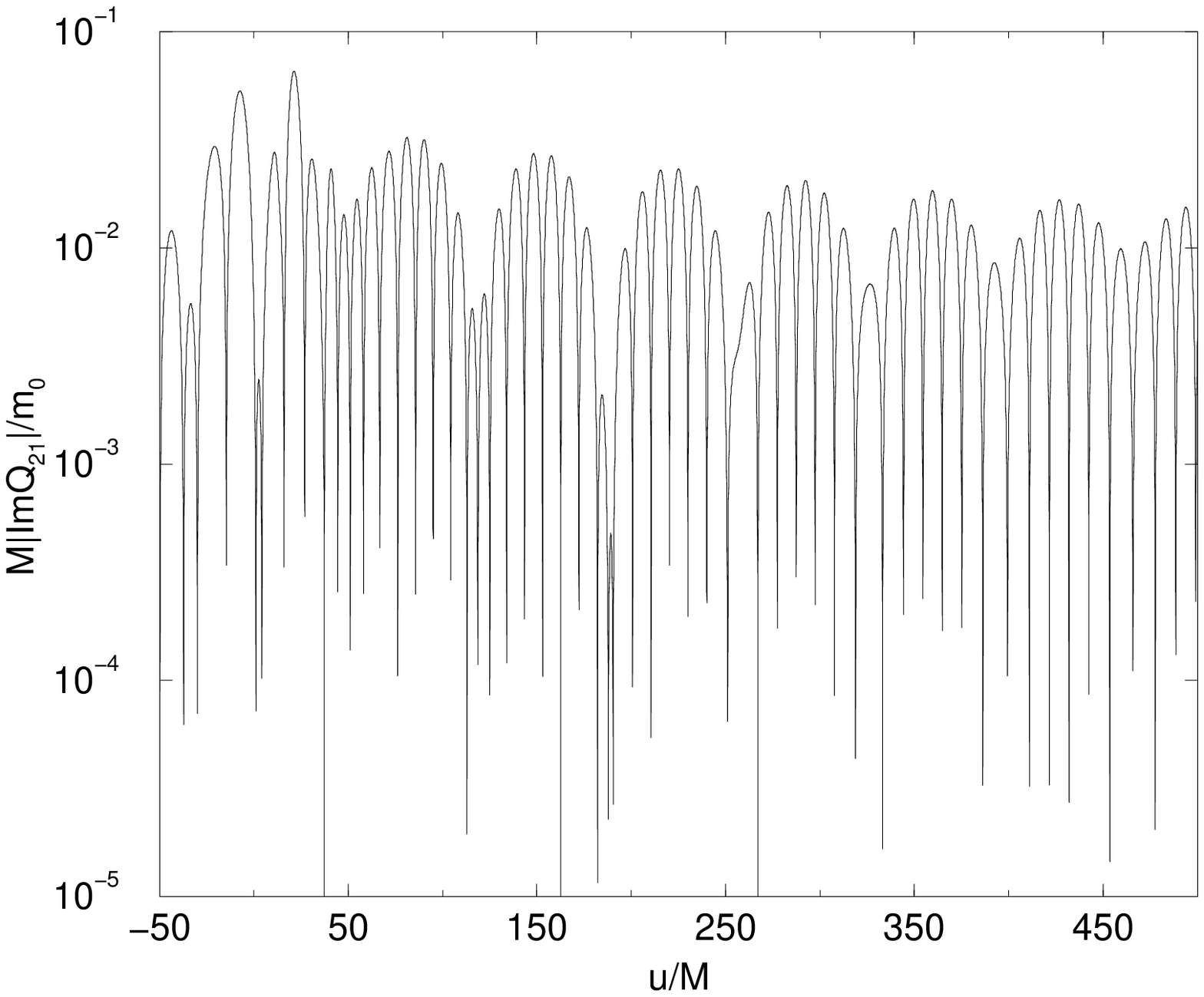}\hspace*{.1\textwidth}
\epsfxsize=0.35\textwidth \epsfbox{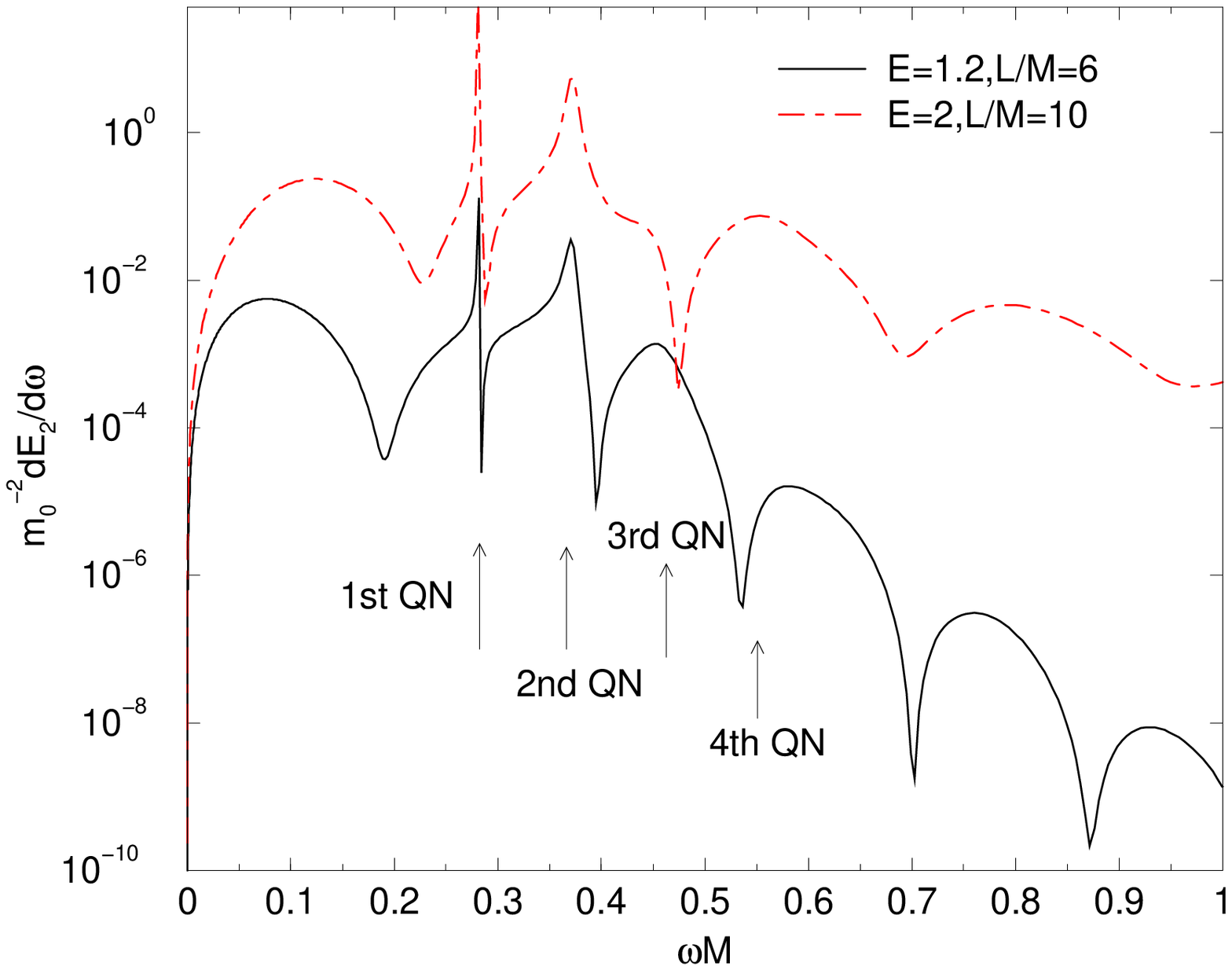}
\caption{\label{HomogstarR=2.33wformspect} Gravitational radiation
from a particle scattered by a homogeneous star of radius $R=2.33M$. (a) Absolute value of 
the imaginary part of $Q_{21}$ in a linear-log scale for a particle with 
parameters $E=2,L=10M$ and turning point $10M/3$. (b) The energy 
spectra for the same particle and for a particle with a lower energy 
($E=1.2,L=6M$ and turning point $r_{t}=5.163M$). The four lowest damped 
$l=2$ QN modes, with frequencies $\omega_{I}M=0.281+i0.00014$, 
$\omega_{II}M=0.372+i0.0048$, $\omega_{III}M=0.458+i0.025$ and 
$\omega_{IV}M=0.554+i0.049$ are indicated by arrows.}
\end{figure}

Despite this element of angular choice, the waveforms are the
preferred tools for assessing the excitation of QN ringing in cases,
like that of Fig.~\ref{BHspect}, for which the QN mode ``resonance''
is broad.  In the case of narrow resonances, such as the QN modes for
ultracompact (and probably unrealistic) stellar models, the spectrum
itself is the better indicator. This is clearly shown in
Fig.~\ref{HomogstarR=2.33wformspect} in which results are given for a
constant density stellar model with $R=2.33M$, very close to the
lowest allowed limit $R/M=9/4$.  For this stellar model the four
lowest damped $l=2$ QN modes are $\omega_{I}M=0.281+i0.00014$,
$\omega_{II}M=0.372+i0.0048$, $\omega_{III}M=0.458+i0.025$ and
$\omega_{IV}M=0.554+i0.049$.  The results are given for $l=2$
radiation from a particle with $E=2,L=10M$.  Plots are presented for
the waveform, in the direction $\theta=\pi/2, \varphi=0$, and for the
energy spectrum.  It is apparent in the latter that the first two QN
modes of the star are excited and there are hints that all the first
four might also be.  The picture given by the waveform is rather more
confusing due to the simultaneous presence of the two modes
$\omega_{I}$ and $\omega_{II}$ with very small damping. The most
obvious feature of the waveform, in fact is the beat frequency between
these two almost periodic modes.  For the case of this stellar model
with several competing QN frequencies, and with sharp QN resonances,
the nature of QN excitation is more clearly shown by the spectrum than
the waveform. A second spectrum in
Fig.~\ref{HomogstarR=2.33wformspect}, for a particle with a smaller
energy of $E=1.2$ shows that the spectrum is equally useful for a
particle orbit that is not highly relativistic.  An interesting
feature of the spectra is that the $w$ modes appear as narrow spikes
at the QN resonances. These spikes contain little energy, but they
appear at the high frequency boundaries of broad spectral peaks very
distinct from the particle radiation spectrum.

\begin{figure}
\hspace*{.1\textwidth}\epsfxsize=0.35\textwidth
\epsfbox{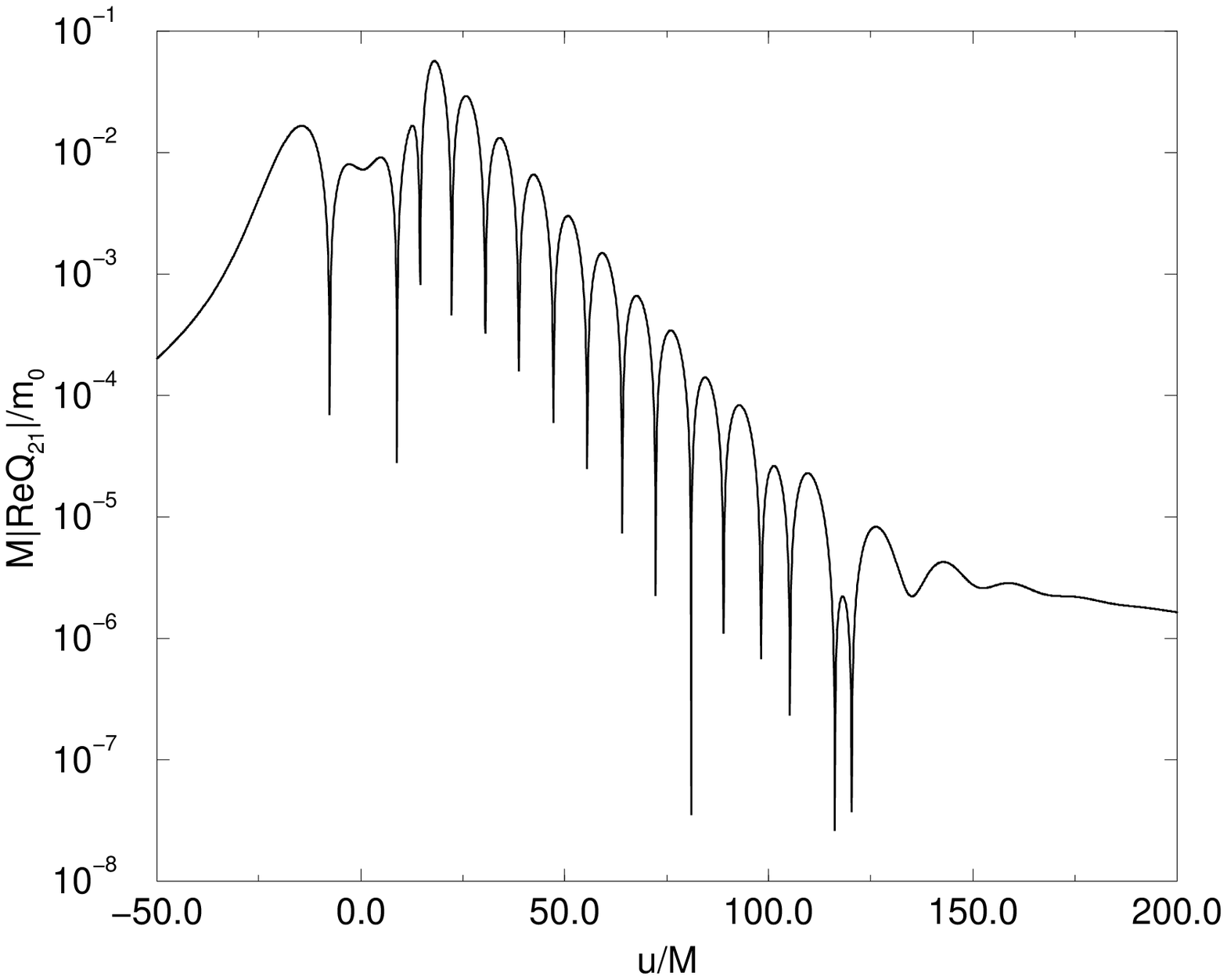}\hspace*{.1\textwidth} \epsfxsize=0.35\textwidth
\epsfbox{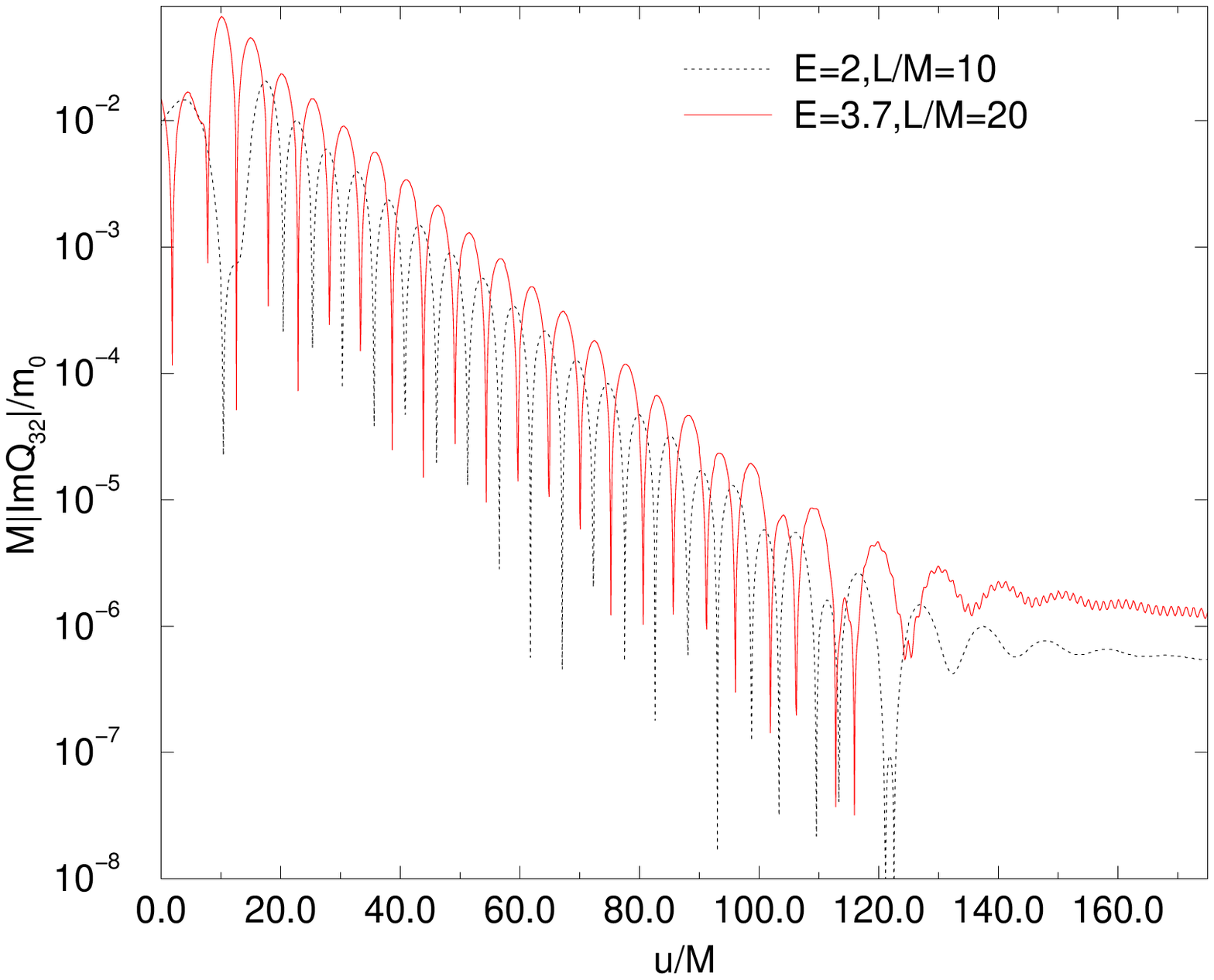} 
\caption{\label{BHwformlog} In (a) the absolute value of the $l=2$ 
waveform of
figure 1.b is plotted in a linear-log display.
The $l=3$
waveform in the plane $\theta=\pi/2$ and direction $\varphi=0$, is shown 
in (b) along with the $l=3$ waveform for an orbit
with $E=3.7$, $L=20M$, and $r_{t}= 3.768M$.
}
\end{figure}

The very sharp QN resonances appear to be unique to ultracompact
stellar models.  Since we will be concerned here, for the most part,
with more-or-less realistic models, with broad QN resonances, we will
rely on waveform pictures in our study of excitation of QN ringing. We
will choose to clarify the presence of QN ringing in these waveforms
by displaying them as linear-log plots of the magnitude of the
waveform. In such plots QN ringing appears as a distinct sequence of
bumps whose peaks lie on a straight line.  The spacing of the bumps
gives the real part of the QN frequency and the slope of the line
gives the imaginary part. The distinctive nature of QN ringing in such
a plot is shown in Fig.~\ref{BHwformlog}.  The waveform is shown in
(a) and (b) respectively for the $l=2$ and $l=3$ multipoles of the
orbit used for Fig.~\ref{BHspect}. In (b) the waveform is also shown
for an orbit with $E=3.7$, $L=20M$, and a turning point at $r_{t}=
3.767501M$.  The pattern of bumps in part (b) has the same spacing and
slope for both orbits, confirming the fact that the QN frequency is
independent of the source. The higher oscillation rate and roughly
equal damping for $l=3$ in comparison with $l=2$ is evident in the
patterns in (a) and (b). The QN frequencies found from these
linear-log plots are $\omega M=0.374+i0.090$ for $l=2$, and $\omega
M=0.600+i0.093$ for $l=3$. These are in excellent agreement with the
known values $\omega M=0.3736+i0.089$ for $l=2$, and $\omega
M=0.5994+i0.0927$ for $l=3$. The combination of parameters ($E=3.7,
L=20M$) for the comparison waveform in Fig.~\ref{BHwformlog}\,b was
chosen because it gives excitation of QN ringing comparable to that
for $E=2, L=10M$. It is worth noting that the particle frequency
$2d\varphi/dt$ is almost identical for the two sets of parameters.

\begin{figure}
\hspace*{.1\textwidth}\epsfxsize=0.35\textwidth
\epsfbox{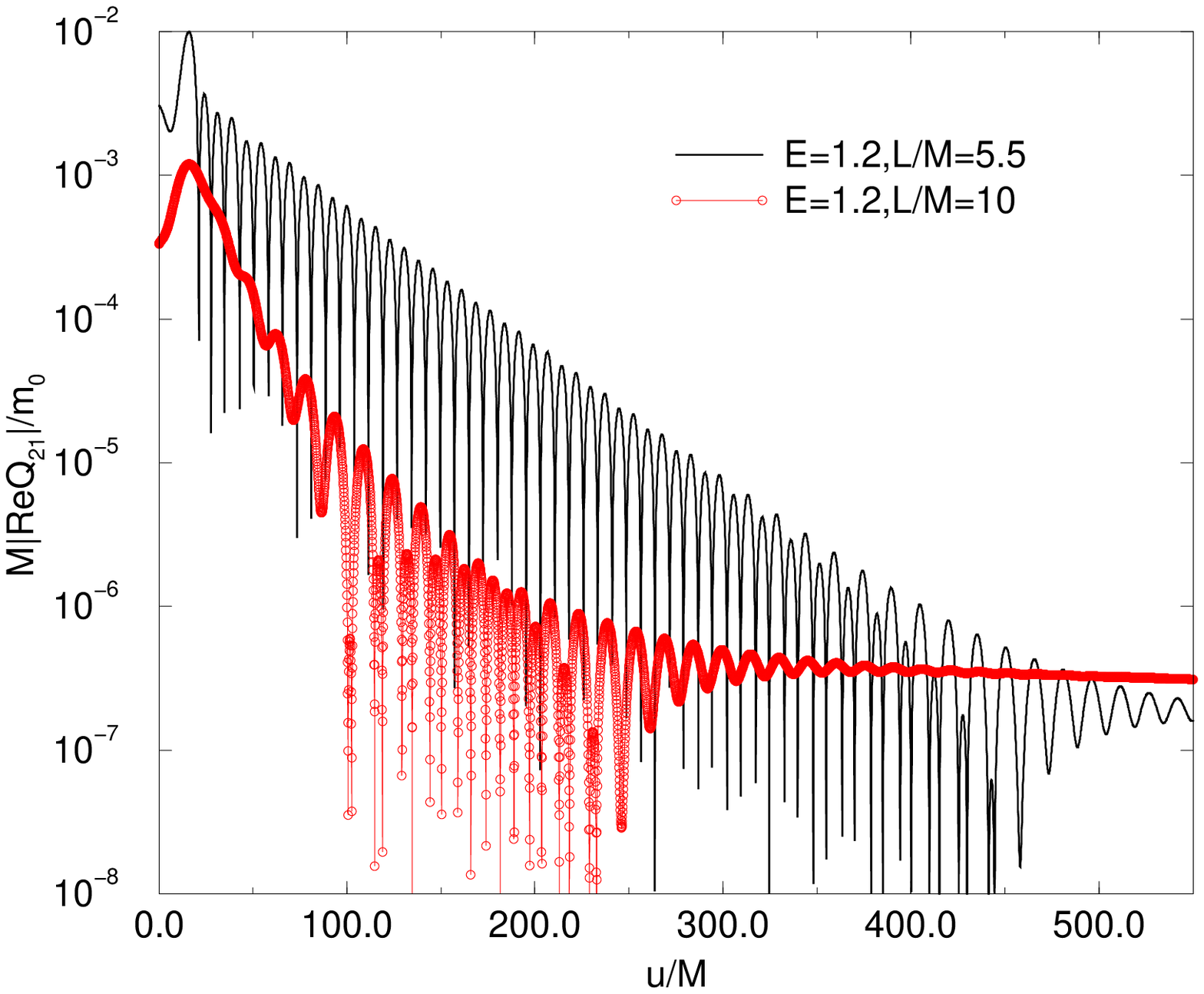}\hspace*{.1\textwidth} \epsfxsize=0.35\textwidth
\epsfbox{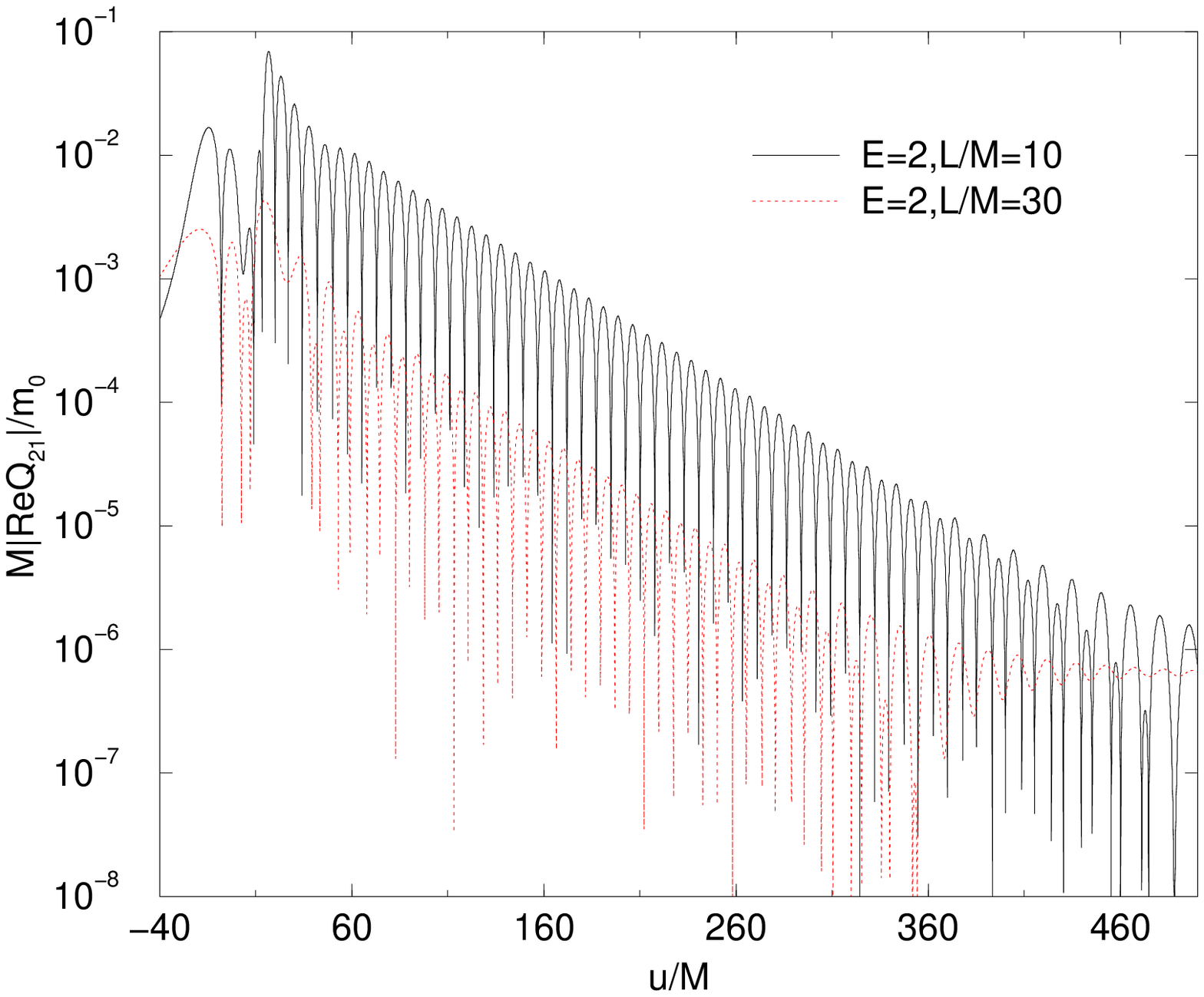} 
\caption{\label{HomogstarR=2.5wformlog} 
The real part of the $l=2, m=1$ odd parity waveform $Q_{21}$, for a
constant density star with $R=2.5M$.
The waveforms in (a) are for two orbits with $E=1.2$.
The low $L$ orbit has $L=5.5M$, $r_{t}=4.047M$, and a particle
frequency $d\varphi/dt=0.142M^{-1}$; 
the high $L$ orbit has $L=10M$, $r_{t}=11.633M$, and a particle
frequency $d\varphi/dt=0.051M^{-1}$. The waveforms in (b) are for
two orbits with $E=2$.
The low $L$ orbit has $L=10M$, $r_{t}=10M/3$, and a particle
frequency $d\varphi/dt=0.180M^{-1}$; 
the high $L$ orbit has $L=30M$, $r_{t}=15.862M$, and a particle
frequency $d\varphi/dt=0.052M^{-1}$. The least damped ($l=2$) odd parity QN mode of this star has frequency $\omega M=0.412+i0.0219$.}
\end{figure}

In Fig.~\ref{HomogstarR=2.5wformlog} we present linear-log waveforms
for a constant density stellar model with $R/M=2.5$, an ultracompact
stellar model that is astrophysically implausible, but that is useful
as an illustration.  The real part of the $l=2, m=1$ waveforms are
shown for particle orbits with the relatively low energy $E=1.2$, and
for a higher energy $E=2$. For each energy one model is a low $L$
orbit with a small turning point $r_{t}$, and the other model has
large $L$ and large $r_{t}$. In each case the large $r_{t}$ orbit
gives far weaker QN ringing than does the small $r_{t}$ orbit.  The
long sequence of well defined bumps for the small $r_{t}$ models in
Fig.~\ref{HomogstarR=2.5wformlog} allow a graphical determination of
the least damped $l=2$ QN frequency $\omega M=0.4123+i0.02199$, to
accuracy of better than 1\%.

\begin{figure}
\hspace*{.1\textwidth}\epsfxsize=0.35\textwidth
\epsfbox{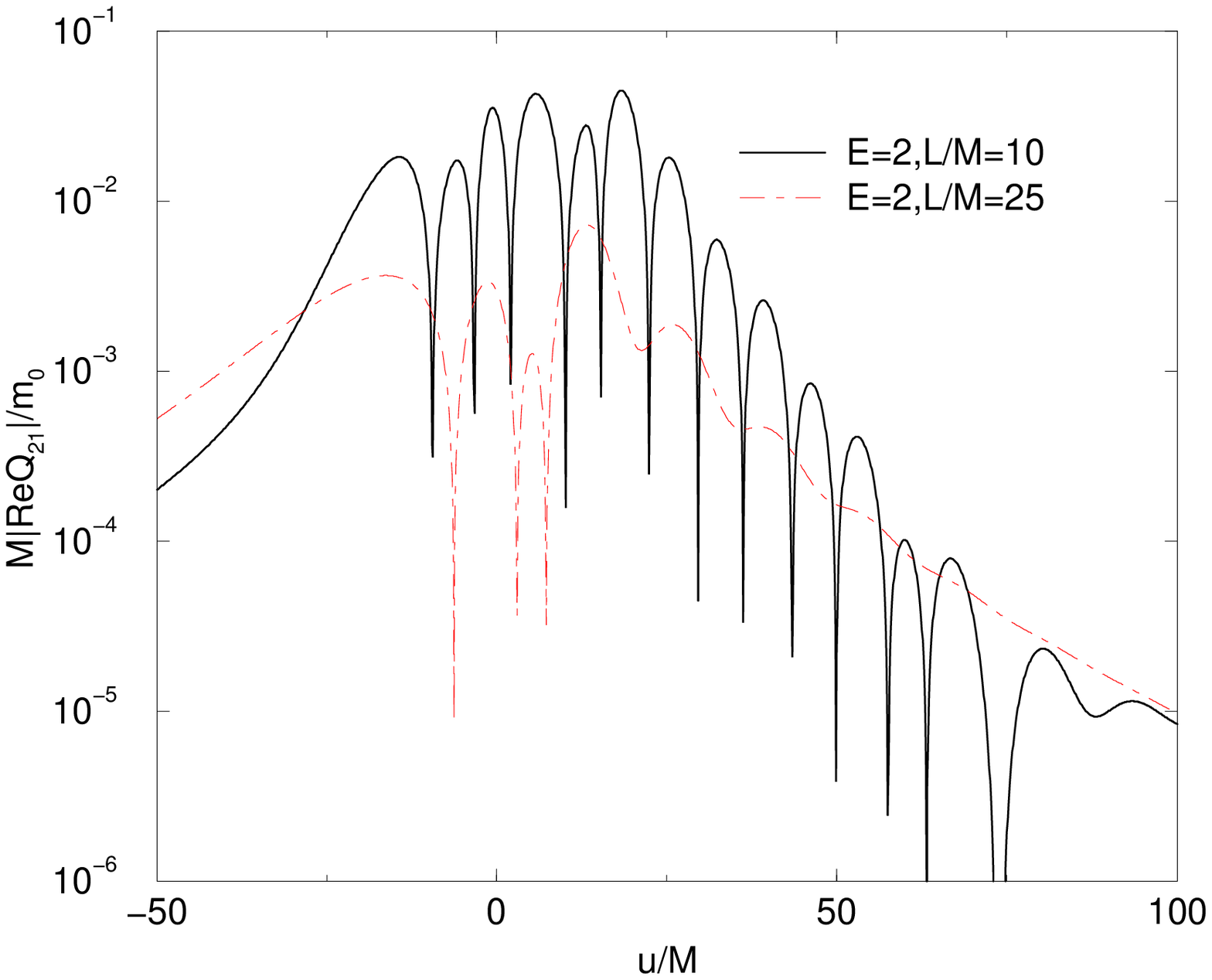}\hspace*{.1\textwidth} \epsfxsize=0.35\textwidth
\epsfbox{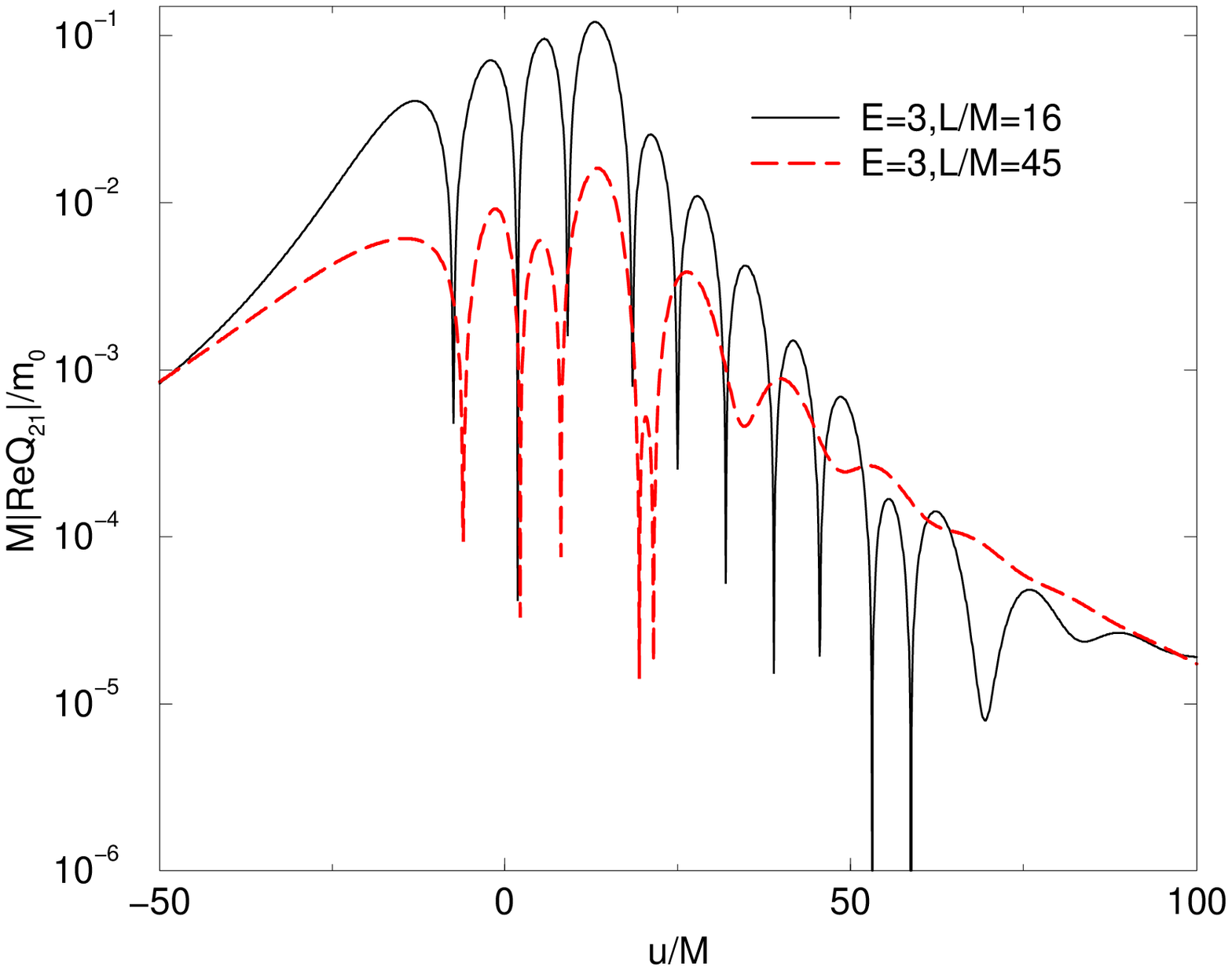} 
\caption{\label{HomogstarR=3.125wformlog} 
The real part of the $l=2, m=1$ odd parity waveform $Q_{21}$, for a
constant density star with $R=3.125M$.  The waveforms in (a) are for
two orbits with $E=2$.  The waveform with pronounced QN ringing has
$L=10M$, $r_{t}=10M/3$ and a particle frequency
$d\varphi/dt=0.180M^{-1}$; the waveform with weak QN ringing has $L=25M$,
$r_{t}= 12.94M$ and $d\varphi/dt=0.063M^{-1}$.  The waveforms in (b) are
for particle orbits with $E=3$.  The waveform with strong QN ringing
corresponds to a particle orbit with $L=16M$, $r_{t}= 3.729M$ and
$d\varphi/dt=0.178M^{-1}$; the waveform with weak ringing corresponds to a
particle orbit with $L=45M$, $r_{t}= 14.66M$ and
$d\varphi/dt=0.0603M^{-1}$.
The strong ringing part can be fitted with precision better than 1\% by 
an oscillating damped function with the frequency of the lowest $l=2$ QN mode 
which is $\omega M=0.455+i0.1386$.
}
\end{figure}

Results for orbits around a constant density model with $R/M=3.125$
are shown in Fig.~\ref{HomogstarR=3.125wformlog}, for two different
values of $L$ at each of two different values of $E$. For this model
the least damped $l=2$ QN mode has a frequency $\omega
M=0.455+i0.1386$ with a real part slightly higher than for a black
hole ( $0.39M^{-1}$) or an ultracompact $R/M=2.5$ model
($0.41M^{-1}$). More important, the imaginary part of the QN frequency
is large, around 50\% larger than for a black hole and more than 6
times that for the ultracompact stellar model. 
The strong ringing part observed for the two smaller angular momenta can 
be approximated with precision better than 1\% by 
the lowest damped QN mode.

\begin{figure}
\hspace*{.1\textwidth}\epsfxsize=0.35\textwidth
\epsfbox{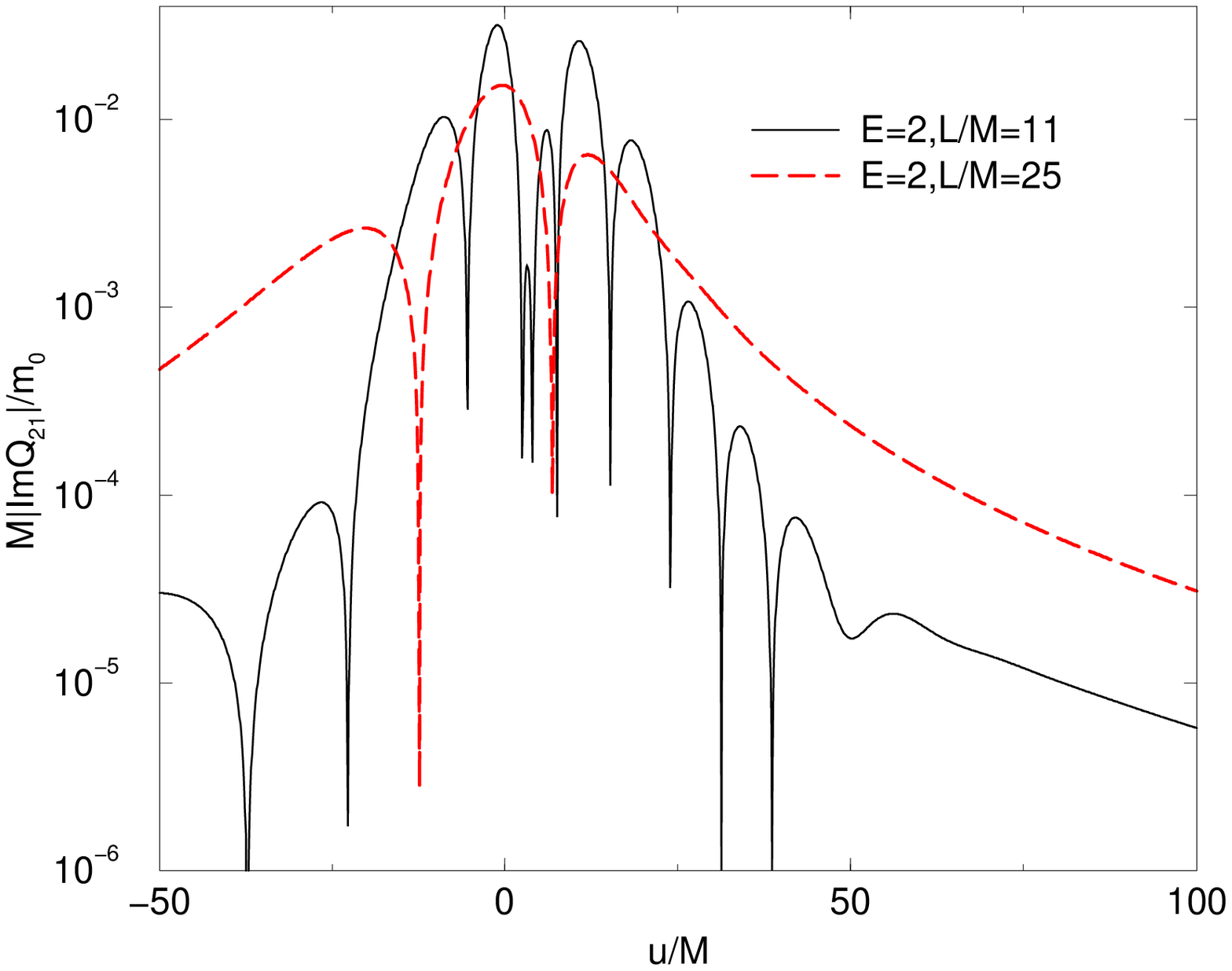}\hspace*{.1\textwidth}
\epsfxsize=0.35\textwidth \epsfbox{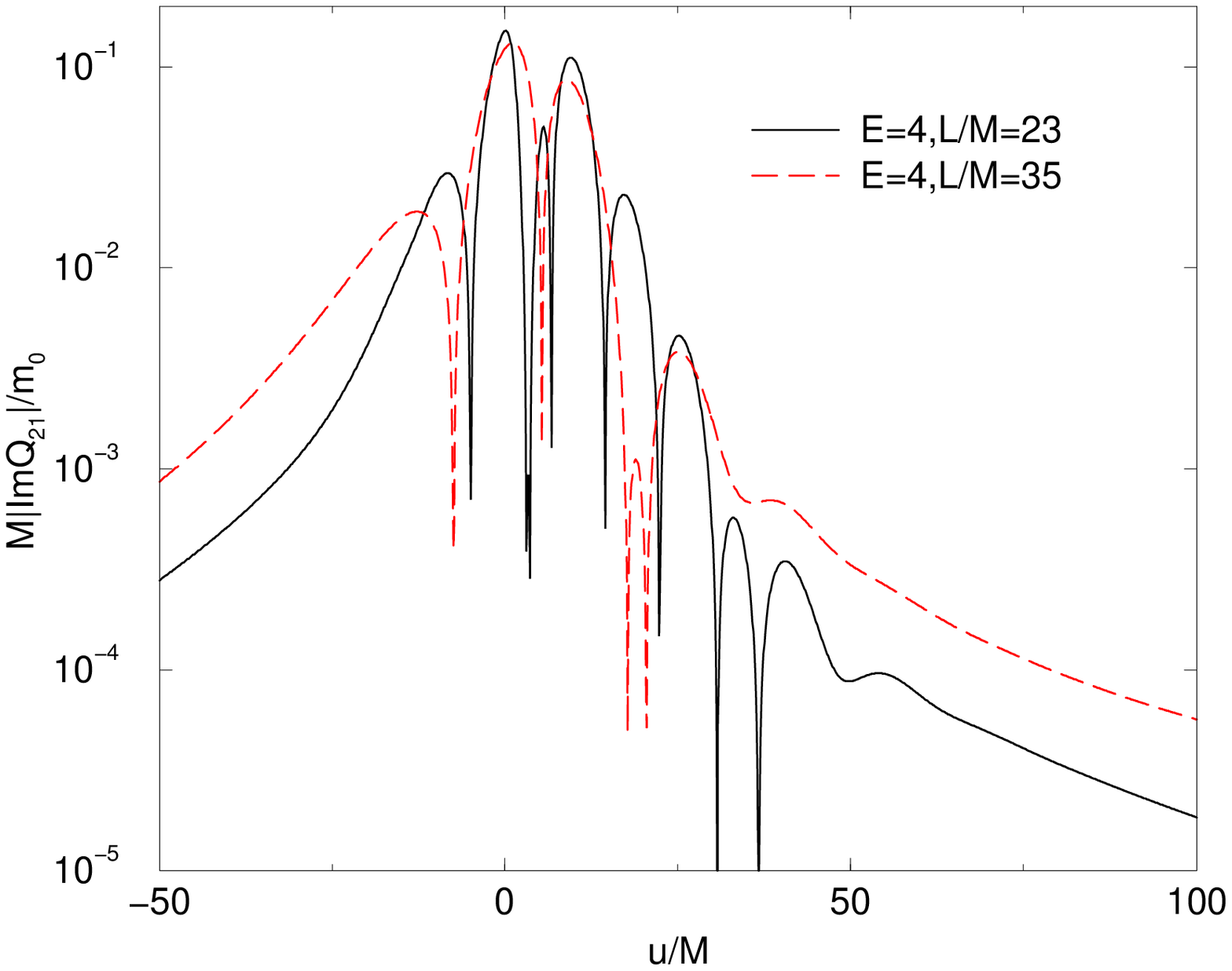}
\caption{\label{HomogstarR=4wformlog} The real part of the $l=2, m=1$
odd parity waveform $Q_{21}$, for a constant density star with $R=4M$.
The waveforms in (a) are for two orbits with $E=2$.  The waveform with
pronounced QN ringing has $L=11M$, $r_{t}= 4.34M$ and a particle
frequency $d\varphi/dt=0.157M^{-1}$; the waveform with no evidence of QN
ringing has $L=25M$, $r_{t}=12.94M$ and $d\varphi/dt=0.0631M^{-1}$.  The
waveforms in (b) are for particle orbits with $E=4$.  The waveform
with strong QN ringing corresponds to a particle orbit with $L=23M$,
$r_{t}=4.259M$ and $d\varphi/dt=0.1682M^{-1}$; the waveform with weak
ringing corresponds to a particle orbit with $L=35M$, $r_{t}= 7.71M$
and $d\varphi/dt=0.109M^{-1}$.
For a constant density star with $R=4$
the least damped odd parity ($l=2$) QN mode has frequency $\omega M=0.401+i0.209$.
}
\end{figure}

\begin{figure}
\hspace*{.1\textwidth}\epsfxsize=0.35\textwidth
\epsfbox{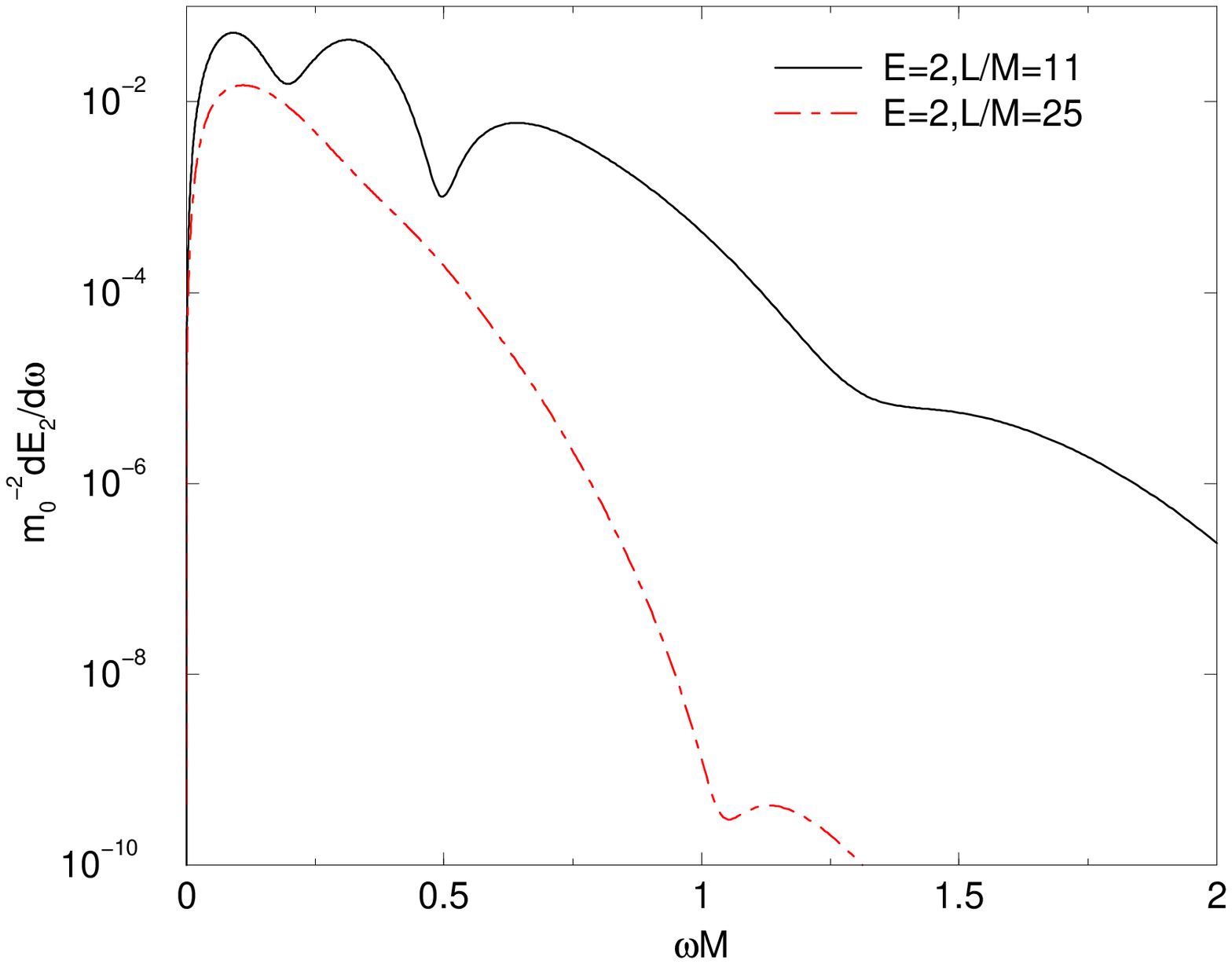}\hspace*{.1\textwidth}
\epsfxsize=0.35\textwidth \epsfbox{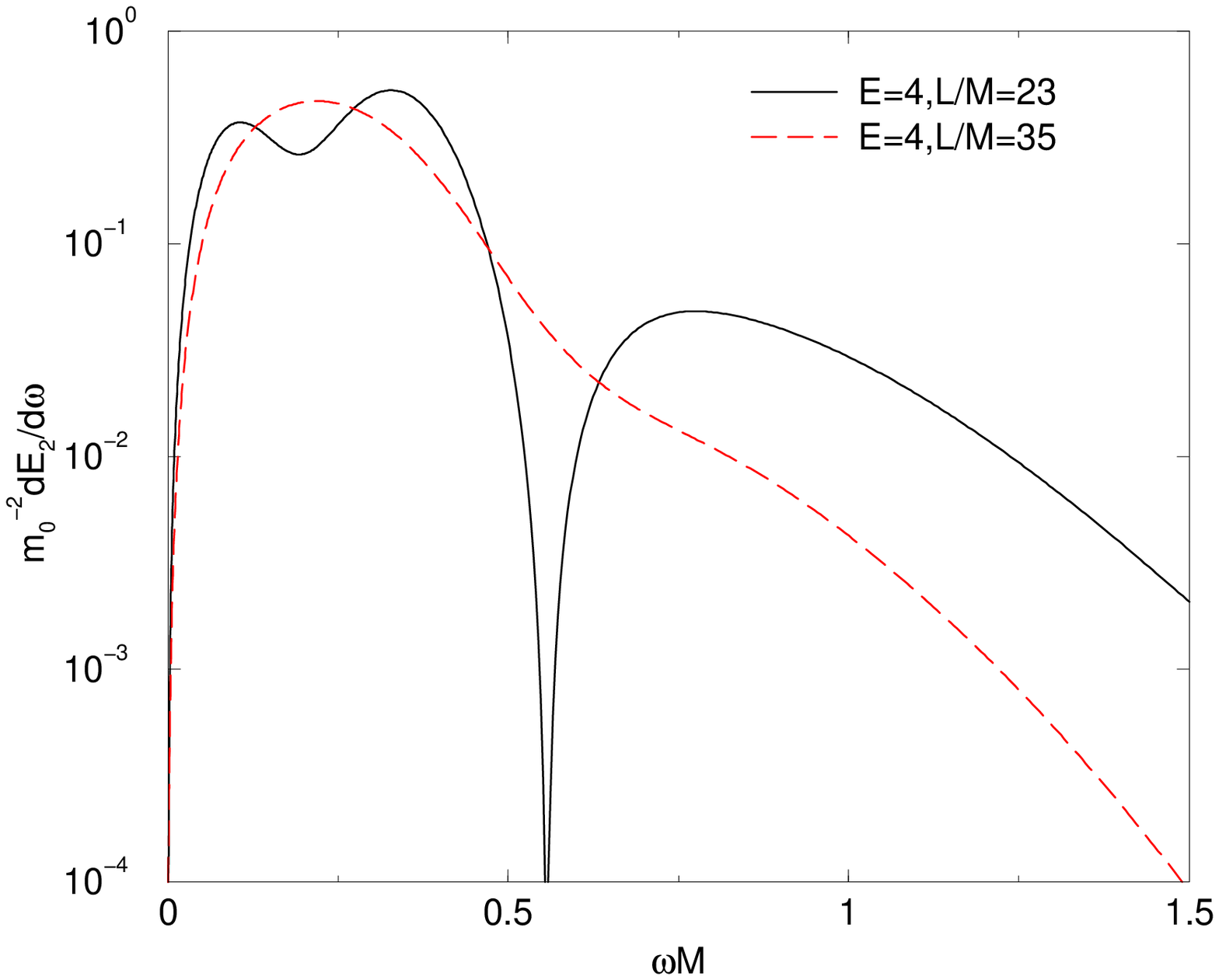}
\caption{\label{HomogstarR=4spectra} The energy spectra for the case
depicted in the previous figure. The background spacetime is a
homogeneous star of radius $R=4M$.  }
\end{figure}

The patterns and trends in Figs.~\ref{HomogstarR=2.5wformlog} and
\ref{HomogstarR=3.125wformlog} continue in
Fig.~\ref{HomogstarR=4wformlog}, in which waveforms are given for
constant density models with $R/M=4$, a value close to the typical
radius of a neutron star. In this case the least damped $l=2$ QN mode
has frequency $\omega M=0.401+i0.209$, so the QN resonance is
extremely broad. A consequence is that the pattern of QN bumps is less
clear in the linear-log waveforms. For the low $L$, high particle
frequency orbits, there is evidence of excitation of the lowest damped 
QN mode for
both energies, but none for the high values of L. The energy spectra
for these models are presented in Fig.~\ref{HomogstarR=4spectra}.  For
homogeneous stars with $R/M>4$, the damping of QN modes will be even
stronger than in Fig.~\ref{HomogstarR=4wformlog}, and QN ringing will
be even less apparent in the waveform.

\begin{figure}
\hspace*{.1\textwidth}\epsfxsize=0.35\textwidth
\epsfbox{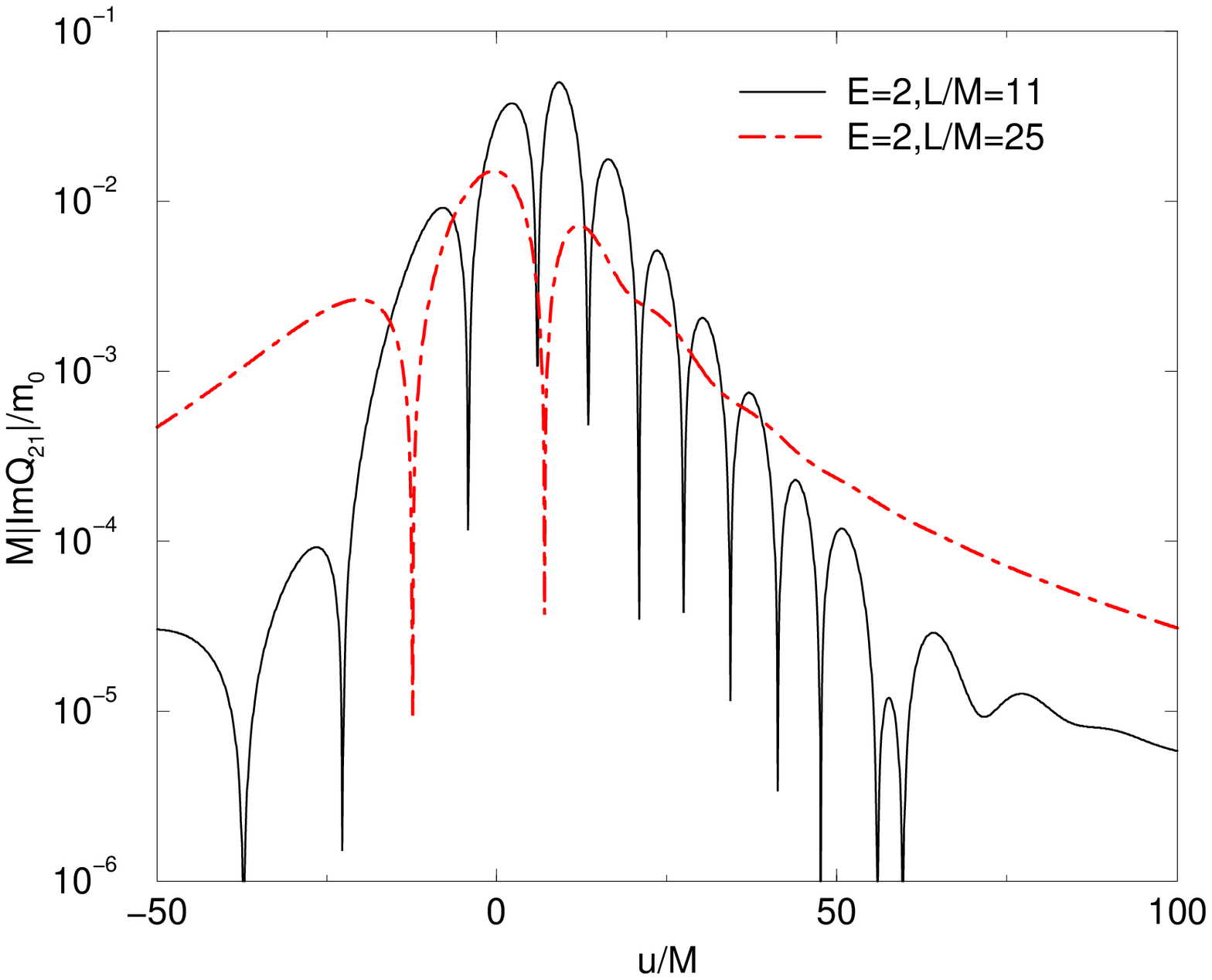}\hspace*{.1\textwidth}
\epsfxsize=0.35\textwidth \epsfbox{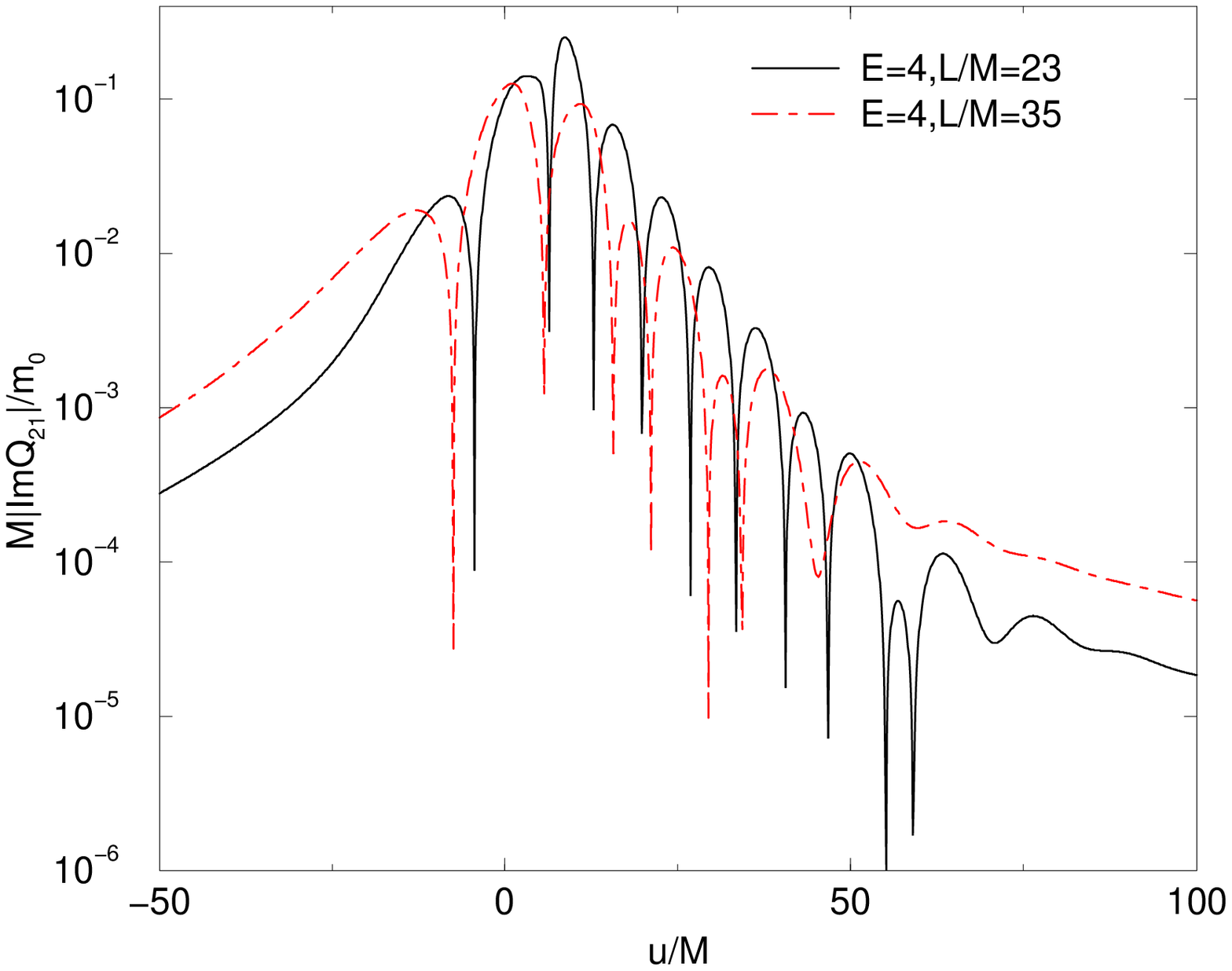}
\caption{\label{Polytropic}The absolute value of the imaginary part of the 
odd parity quadrupole waveform $Q_{21}$ is shown for a polytropic star of 
index $n=1$ and $K=100 Km^{2}$. The central density of this star, 
$\rho_{0}=5\times 10^{15}g/cm^{3}$ and the radius is $R=3.91M$. In (a) the 
scattered particle has energy $E=2$ and two angular momenta, $L=11M$ 
(turning point 4.344M,$d\varphi/dt=0.157M^{-1}$) and $L=25M$ (turning point 
$12.943M$,$d\varphi/dt=0.063M^{-1}$) are considered. In (b), a higher energy 
($E=4$) is considered and also the waveform for two angular momenta are shown: 
$L=23M$ (turning point 4.259M,$d\varphi/dt=0.168M^{-1}$) and $L=35M$ 
(turning point is 7.71M,$d\varphi/dt=0.109M^{-1}$).}
\end{figure}

The results presented in the previous figures 
are not sensitively dependent on the assumption of a homogeneous stellar
model. Studies of polytropes with smaller stiffness show the same
pattern of QN excitation as the $n=0$ constant density polytrope. A
typical case is shown in Fig.~\ref{Polytropic}, which displays the
waveform when a particle with energies $E=2,4$ is scattered by a
polytropic star of index $n=1$ and with central density
$\rho_{0}=5\times 10^{15}$\,g/cm$^{3}$. (In the equation of state
$p=KG\rho^{2}$, the parameter $K$ is $10^{8}\,{\rm m}^{2}$). For this
model the radius of the star is $R=3.911M$, and the lowest damped
$l=2$ QN mode has the frequency $\omega M=0.4603+i0.1504$. This mode
is excited by both angular momenta considered in
Fig.~\ref{Polytropic}b where the particle has a very relativistic
energy, while it is only excited by the lowest angular momenta in
Fig.~\ref{Polytropic}a.  From the graphs, the ringing frequency is
computed to be $\omega=0.458+i0.1503$, very close to the QN frequency.

\begin{figure}
\hspace*{.1\textwidth}\epsfxsize=0.35\textwidth
\epsfbox{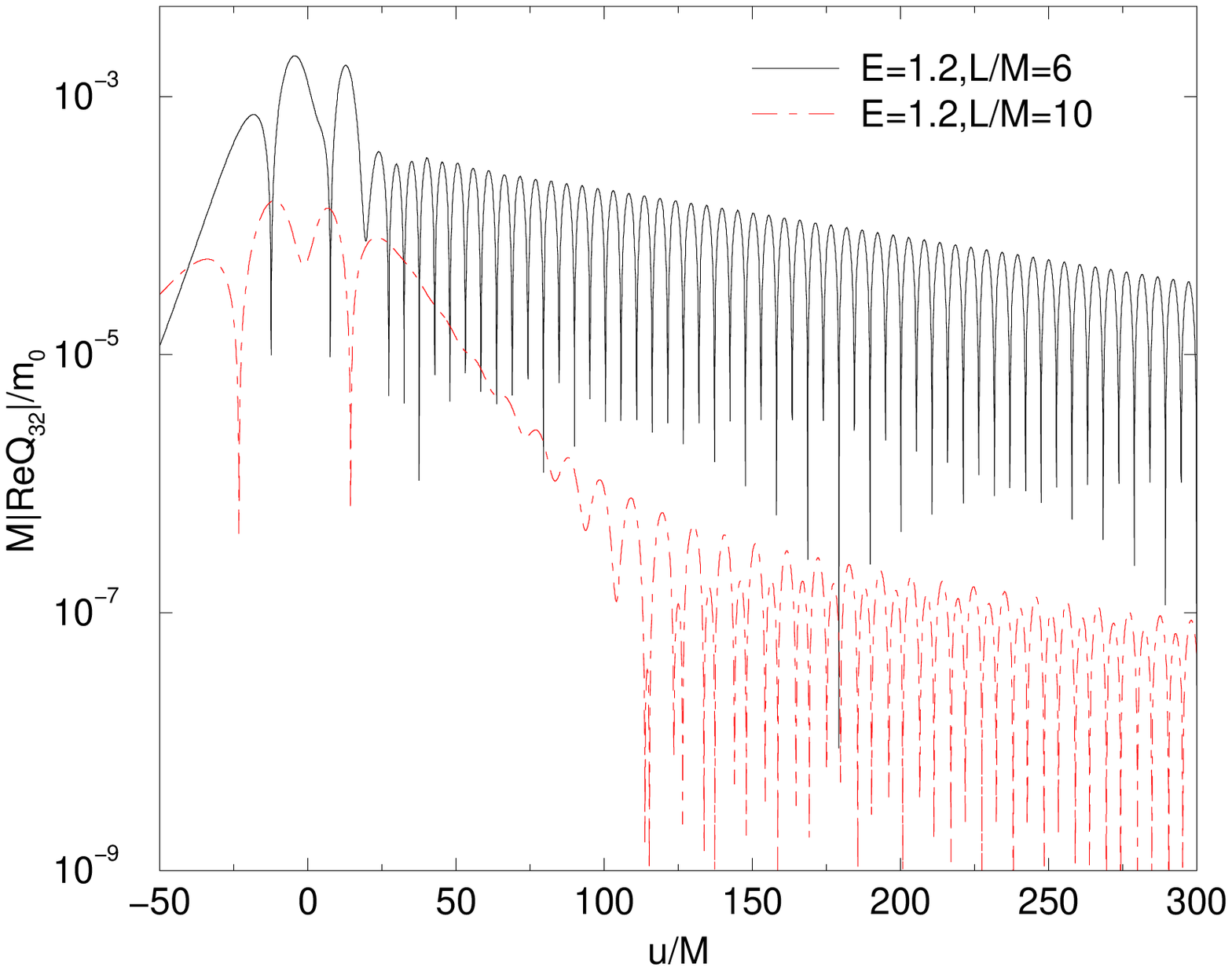}\hspace*{.1\textwidth}
\epsfxsize=0.35\textwidth \epsfbox{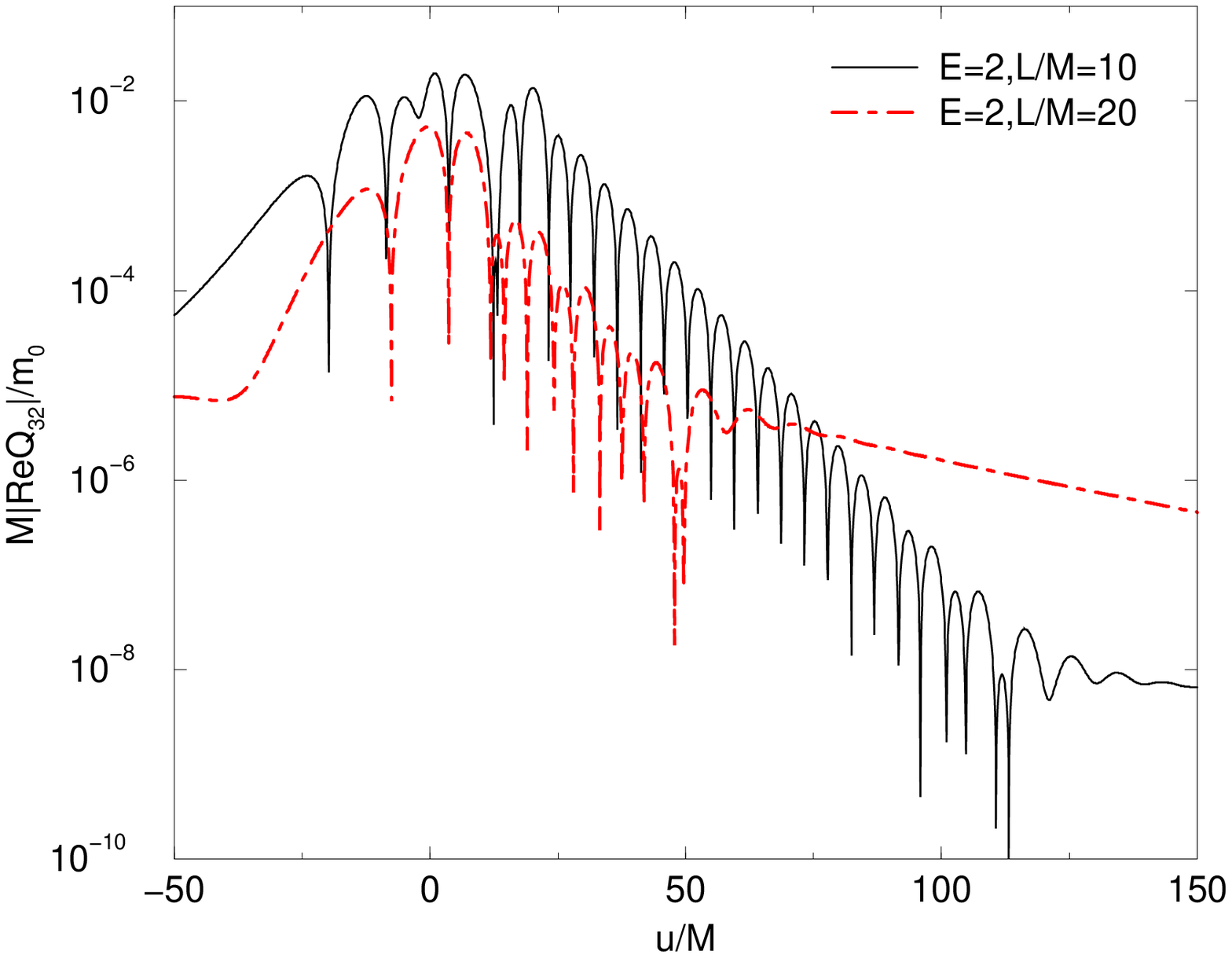}
\caption{\label{HomogstarR=2.5nd3.125wformlogoct}
The absolute value of the real part of the odd parity octupole
waveform $Q_{32}$ is shown for a homogeneous star of radius $R=2.5M$
in (a) and of radius $R=3.125M$ in (b).  In (a) we consider a particle
with small energy ($E=1.2$) and two angular momenta: $L=6M$ (turning
point is 5.163M,$2d\varphi/dt=0.23M^{-1}$) and $L=10M$ (turning
point is 11.633M, $2d\varphi/dt=0.102M^{-1}$). The amplitude of
the wave for the far orbit is so small that the ringing observed (at
about twice the frequency of the QN mode) is probably numerical
error. In (b) we consider a particle with energy ($E=2$) and two
angular momenta: $L=10M$ (turning point is
10M/3M,$2d\varphi/dt=0.36M^{-1}$) and $L=20M$ (turning point is
10M,$2d\varphi/dt=0.16M^{-1}$) moving outside a star with radius
$R=3.125M$. There is, for both angular momenta, excitation of a QN
mode of the star, that most likely is the lowest damped $l=3$ odd mode
(from the graphic for the smaller angular momentum orbit we get $\omega
M=0.691+i0.143$). }
\end{figure}
\begin{figure}
\hspace*{.1\textwidth}\epsfxsize=0.35\textwidth
\epsfbox{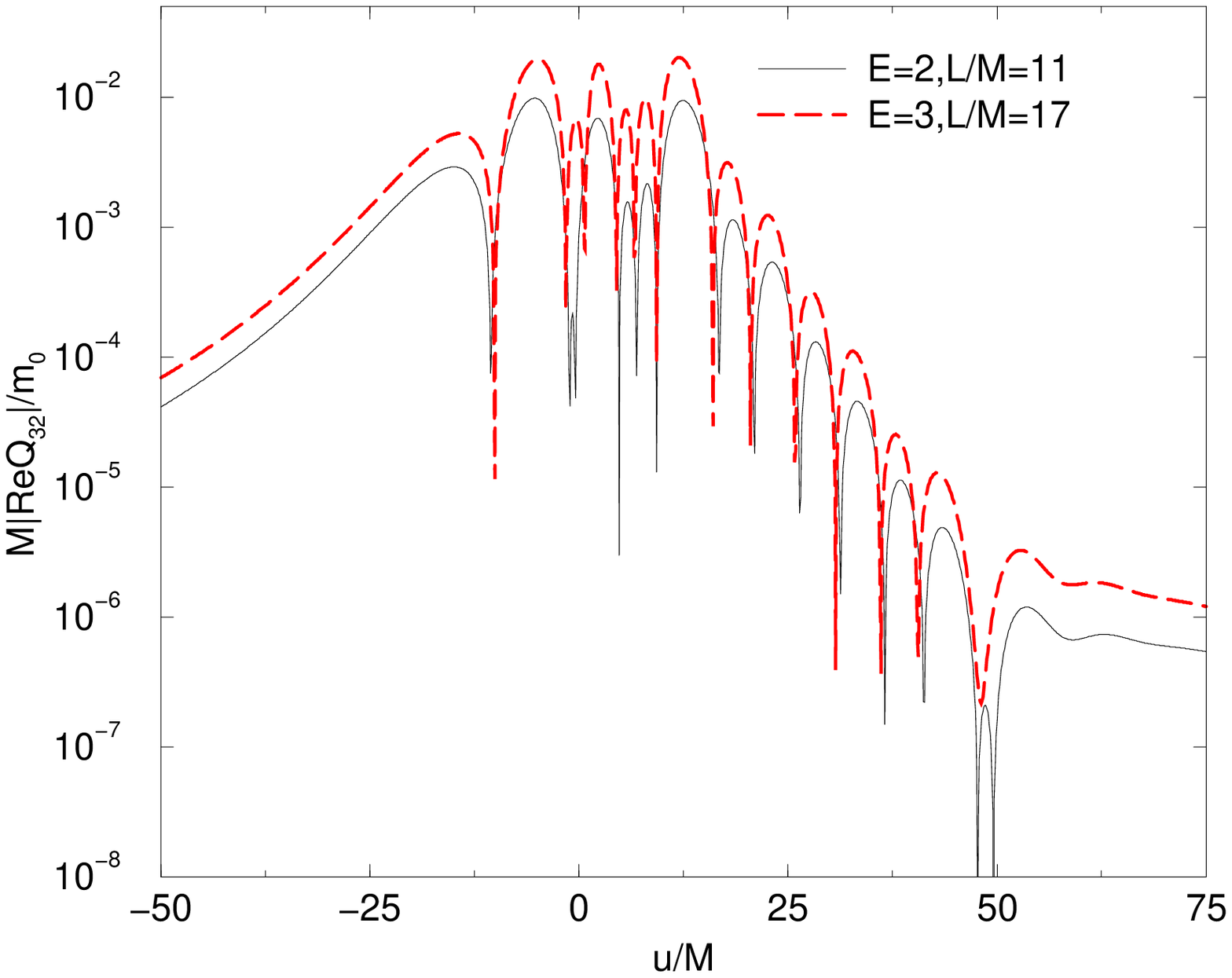}\hspace*{.1\textwidth}
\epsfxsize=0.35\textwidth \epsfbox{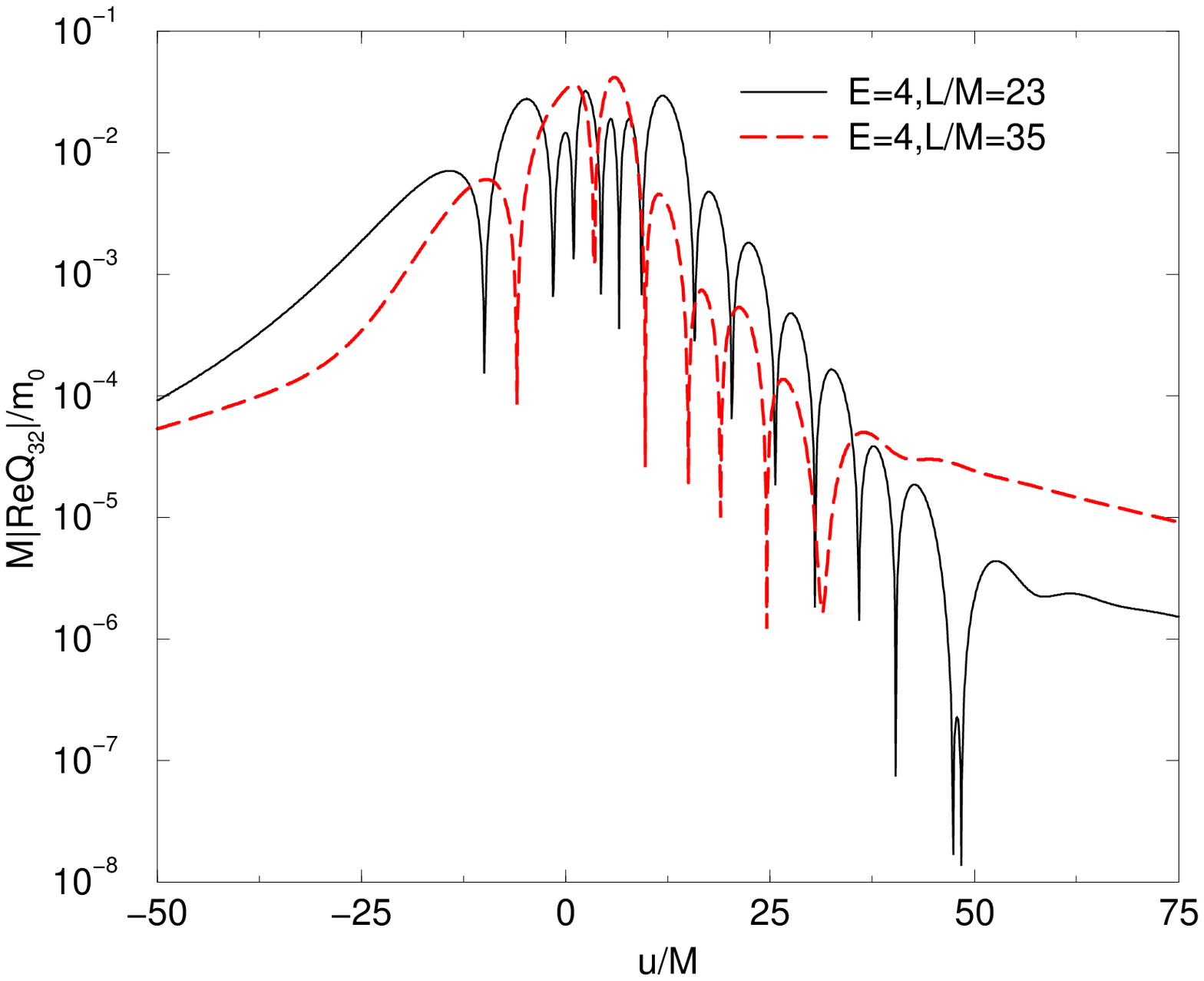}
\caption{\label{HomogstarR=4wformlogoct}The absolute value of the real part of the odd parity octupole waveform $Q_{32}$ is shown for an homogeneous star of radius $R=4M$. In (a) we consider two sets of parameters for the particle: $E=2,L=11M$ (turning point is 4.343M, $2d\varphi/dt=0.315M^{-1}$) and $E=3,L=17M$ (turning point at 4.25M, $2d\varphi/dt=0.333M^{-1}$). In both cases ringing at the same frequency is visible (that frequency is approximately $\omega=0.64+i0.23$.
In (b) the particle has energy $E=4$ and we consider two angular momenta: $L=23$ (turning point $r_{t}=4.259M$,$2d\varphi/dt=0.336M^{-1}$) and $L=35M$ (turning point 7.71M, $2d\varphi/dt=0.22M^{-1}$). In both cases there is ringing. The oscillation is clear enough for the smaller turning point to allow us to obtain its complex frequency: $\omega M=0.64+i0.23$.}

\end{figure}

So far we discussed only excitation of $l=2$ QN modes of stars, but as
we have shown (see Figs.~\ref{BHspect} and \ref{BHwformlog}b), when
the $l=2$ QN mode of a black hole is excited by a scattered particle,
the least damped $l=3$ QN mode is  excited with equal strength. To
investigate whether this is also true  for stellar models, we computed
the octupole waveforms for a small set of homogeneous stars, with
radii $R=2.5M$, $3.125M$ and $R=4M$. The results, displayed in
Figs.~\ref{HomogstarR=2.5nd3.125wformlogoct} and
\ref{HomogstarR=4wformlogoct},  confirm that these compact
stars have their $l=3$ odd parity QN modes excited by scattered particles. 
The $l=3$ results in these figures can be compared with corresponding
$l=2$ results. The orbits for
Fig.~\ref{HomogstarR=2.5nd3.125wformlogoct}a are negligibly different
from the orbits in Fig.~\ref{HomogstarR=2.5wformlog}a. These figures
describe radiation from orbits with the same energy and negligibly
different angular momenta.  The figures show that the $l=3$ QN
frequency, $\omega M=0.598+i0.0087$, has a larger real part than that
of the $l=2$ ringing, but a slower damping time.  The figures show
that there is a similar range of angular momenta for which significant
excitation of the quadrupole and octupole QN modes occurs. One can
also compare Fig.~\ref{HomogstarR=3.125wformlog}a with
Fig.~\ref{HomogstarR=2.5nd3.125wformlogoct}b. In both cases the
particle moves outside an homogeneous star with radius $R=3.125M$ with
the same energy, and again excitation is comparable. In this case, the
$l=3$ QN mode ($\omega M=0.691+i0.143)$ has a damping time almost
equal to the damping time of the $l=2$ QN mode.  A similar comparison
can be made of Fig.~\ref{HomogstarR=4wformlogoct} and
Fig.~\ref{HomogstarR=4wformlog} for stars with $R=4M$.

\section{Conclusions}\label{concl}


QN ringing of $w$ modes make a unique, high frequency, change in the
appearance of waveforms. Even if QN ringing is only weakly excited it
would in principle be easily recognizable and, as pointed out by
Andersson and Kokkotas\cite{AK1}, would provide important information
about the structure of the neutron star emitting the waves.  Our
results above show that $l=3$, and higher multipoles, tend to be
excited on the same order as $l=2$ modes for scattering orbits.  This
is a somewhat unusual finding since gravitational processes tend to be
dominated by quadrupole radiation. Significant higher multipole
radiation may be specific to scattering by relativistic orbits
of a ``particle'' (with negligible angular extent), and
may not be generalizable to more astrophysically plausible sources.
If these higher $l$ modes do turn out to be as highly excited as the
$l=2$ mode, the modes will all be mixed together in the waveform
recorded by a detector; a detector at a single location cannot
``project out'' individual multipole moments as we have done in the
figures presented above.  The signal from a neutron star could then be a
highly complex pattern that in principle contains a great deal of
information about the neutron star. In practice, the signal to noise
ratio, and the breadth of the QN ``lines'' would make it very unlikely
that QN mode spectroscopy could reveal the details of neutron star
structure.  Another practical consideration is that the $w$ mode
frequencies (typically $10^{5}$\,Hz) are well above the upper limit
(several hundred Hz) of laser interferometric detectors, the detectors
that are most capable of detecting waveforms.  The $w$ mode
frequencies, of course, scale inversely as the mass of the neutron
star.  But this is of little consequence, since neutron star masses
are constrained to a very small range. By contrast there is at least
the possibility for black holes of large enough mass (hundreds of solar
masses) that QN ringing will be within the optimal bandwidth of
interferometric detectors.


From the representative waveforms and spectra presented we can infer
that the lowest frequencies of gravitational radiation are negligibly
affected unless QN ringing is very strong, and hence for a laser
interferometric detector, an analysis omitting general relativistic
effects would be satisfactory.  On the other hand, for a detector with
sensitivity extending to hundreds of kilohertz, the $w$ mode
contribution could be very important. The ringing of $w$ modes
provides a high frequency component of the spectrum that could contain
more energy than that due ``directly'' to the particle.  For that
reason it is of some interest to look through the neutron star results
to make inferences about the conditions necessary for $w$ modes to be
strongly excited.  It is tempting to conclude that the excitation of
$w$ modes is largest when the particle frequency is closest to the
real part of the QN frequency.  This conclusion, however, is somewhat
uncertain.  The particle frequencies are always below the QN
frequency, so a {\em higher} particle frequency is the same as a
particle frequency that is {\em closer} to the QN resonance. The most
important determinant of the particle frequency is the turning point
radius $r_{t}$, and smaller $r_{t}$ will produce higher particle
frequency, and hence a particle frequency closer to the QN
resonance. But a smaller $r_{t}$ also means that the source of the
disturbance to the neutron star is closer and hence is better able to
excite QN ringing.

The difference between these two interpretations is important if we
try to generalize from the above results to neutron star processes
that arise more naturally astrophysically. In this connection it is
helpful to turn attention to the (astrophysically implausible)
ultracompact stellar model of Fig.~\ref{HomogstarR=2.33wformspect}
. In the $R/M=2.33$ model the two least damped QN modes have almost
negligible imaginary parts, so the resonance picture would imply that
excitation could be stimulated only by the high frequency tail of the
particle radiation, which is negligible at the QN ``resonant''
frequency (the real part of the QN frequency). In spite of this the
spectrum in Fig.~\ref{HomogstarR=2.33wformspect}b is far from a
particle spectrum. There is in fact more radiation in the high
frequency part of the spectrum than in the low.  This implies that
care must be used in applying the ideas of normal mode systems to QN
modes, and that excitation may be much larger than a simple resonance
model would suggest.

For the models that are astrophysically most relevant, those with
$R/M>4$, the model results suggest that even the source most tightly
coupled to the QN modes (i.e., even the smallest turning point) can
stimulate only moderate ringing. This may be associated with the very
short damping times for these stellar models and may not apply to
stellar models with longer damping times\cite{KJELL,LIND}.  If we
disregard such possibilities, and assume that the QN frequencies of
stellar models are similar to those of polytropic equations of state,
we infer that $w$ mode effects on the radiation from scattering orbits
are not of crucial significance.  We conjecture that supernova core
collapse or the coalescence at the end point of binary inspiral will
couple no better to $w$ modes than do scattering orbits. For $R/M>4$
models, then, $w$ modes will be of crucial importance only if the
source of spacetime perturbations does not stimulate any radiation due
to fluid motion, as is the case for excitation of ringing by an odd
parity gravitational wave.

\section*{acknowledgments}
We thank Nils Andersson for supplying the $l=2$ QN frequencies 
of homogeneous and polytropic stars. We
also thank William Krivan for many useful discussions and for assistance 
in the preparation of the manuscript. ZA was supported by a PRAXIS XXI PhD grant from FCT (Portugal). This work was partially supported
by National Science Foundation grant PHY9734871.

\appendix

\section{Timelike scattered geodesics of Schwarzschild spacetime}
In this appendix we present some of the properties of the family of geodesics of Schwarzschild spacetime considered in this work.  Additional details can be found in \cite{MTW}. 

A particle with a given energy $E$ and angular momentum $L$ (both per unit rest mass), moving in one of these geodesics starts its motion at radial infinity with initial velocity $v_{\infty}=\sqrt{E^{2}-1}/E$,
reaches the turning point $r_{t}$ and returns to infinity. 

Scattered geodesics require $E>=1$. The higher the energy of the particle the more relativistic it is (when $E=1$ the particle falls with zero initial velocity). For a given energy there is a minimum angular momentum if the particle moves in one of these geodesics (see ex. \cite{ON}). It is
\begin{equation}
\frac{L^{2}}{M^{2}}>\frac{27E^{4}-36E^{2}+8+\sqrt{(27E^{4}-36E^{2}+8)^{2}+64(E^{2}-1)}}{2(E^{2}-1)}
\end{equation}
if $E>1$ and
\begin{equation}
\frac{L}{M}>4
\end{equation}
if $E=1$. If the particle has smaller angular momentum and still moves
along a geodesic than it will plunge into the star (or black
hole). The equality happens for a particle moving in a circular
geodesic.

The turning point is the largest of the two positive roots of the cubic equation (the third root is always negative)
\begin{equation}\label{turn}
r^{3}(1-E^{2})-2Mr^{2}+rL^{2}-2ML^{2}=0
\end{equation}
We can find it numerically by noting that it is always located in the region 
\begin{equation}
\frac{L^{2}}{2M}\left(1-\sqrt{1-\frac{12M^{2}}{L^{2}}}\right)<r_{t}<\frac{L^{2}+\sqrt{L^{4}-16M^{2}L^{2}}}{4M}
\end{equation}
and that the other positive root is always located outside this region. For a given energy, the smaller the angular momentum of the particle is, the smaller $r_{t}$ is. However the turning point is always
\begin{equation}
r_{t}>3M
\end{equation}
In the special case $E=1$, (\ref{turn}) becomes a quadratic equation and the turning point can be written in analytical form
\begin{equation}
r_{t}=\frac{L^{2}+\sqrt{L^{4}-16M^{2}L^{2}}}{4M}
\end{equation}
Clearly $r_{t}>4M$ in this case.

Once we know the turning point for a given energy and angular momentum of the particle, we can determine the other two roots of (\ref{turn}) easily. We need first to compute the parameters $e,p$ (see for ex. \cite{CHANDRA}), 
\begin{eqnarray}
& &
e=\frac{L^{2}r_{t}-2Mr^{2}_{t}+L\sqrt{-12L^{2}M^{2}+4L^{2}Mr_{t}+
L^{2}r^{2}_{t}-16M^{2}r^{2}_{t}}}{2M(L^{2}+r^{2}_{t})}\\
& &
p=\frac{r_{t}}{M}(e+1)
\end{eqnarray}
(note that $e=1$ if $E=1$) after what we can rewrite (\ref{turn}) like
\begin{equation}
(r-r_{0})(r-r_{sec})(r-r_{t})=0
\end{equation}
with the three roots given by
\begin{equation}\begin{array}{ccc}
r_{0}=-\frac{pM}{e-1}, & r_{sec}=\frac{2pM}{p-4}, & r_{t}=\frac{pM}{e+1}
\end{array}
\end{equation}

If we choose the origin of coordinate time T, angle $\varphi$ and proper time $\tau$ at the turning point then the value of these quantities at a given value of the radial coordinate $r$ will be determined by the geodesic equations,
\begin{eqnarray}
& &
T(r)=\pm \frac{E}{\sqrt{E^{2}-1}}\int_{r_{t}}^{\infty}\frac{r^{3/2}dr}{(1-2M/r)\sqrt{(r-r_{0})(r-r_{sec})(r-r_{t})}}\\
& &
\varphi(r)=\pm \frac{L}{\sqrt{E^{2}-1}}\int_{r_{t}}^{\infty}\frac{dr}{\sqrt{r(r-r_{0})(r-r_{sec})(r-r_{t})}}\\
& &
\tau(r)=\pm \frac{1}{\sqrt{E^{2}-1}}\int_{r_{t}}^{\infty}\frac{r^{3/2}dr}{\sqrt{(r-r_{0})(r-r_{sec})(r-r_{t})}}
\end{eqnarray}
with the $-$ sign holding for the part of the motion from infinity to $r_{t}$ and the $+$ sign for the part of the motion from $r_{t}$ to infinity.

Although the above integrals are finite, their integrands diverge at the turning point, making difficult their numerical computation. A
possible remedy is to make a change of variable that makes the integrands finite over all the range of integration. For example
\begin{equation}
z=\sqrt{r-r_{t}}
\end{equation}
leads to
\begin{eqnarray}
& &
T(r)=\pm\frac{2E}{\sqrt{E^{2}-1}}\int_{0}^{z(r)}\frac{r^{3/2}dz}{(1-2M/r)\sqrt{(r-r_{0})(r-r_{sec})}}\\
& &
\varphi(r)=\pm\frac{2L}{\sqrt{E^{2}-1}}\int_{0}^{z(r)}\frac{dz}{\sqrt{r(r-r_{0})(r-r_{sec})}}\\
& &
\tau(r)=\pm\frac{2}{\sqrt{E^{2}-1}}\int_{0}^{z(r)}\frac{r^{3/2}dz}{\sqrt{(r-r_{0})(r-r_{sec})}}
\end{eqnarray}
for $E>1$ and to
\begin{eqnarray}
& &
T(r)=\pm\sqrt{\frac{2}{M}}\int_{0}^{z(r)}\frac{r^{3/2}dz}{(1-2M/r)\sqrt{r-r_{sec}}}\\
& &
\varphi(r)=\pm\sqrt{\frac{2}{M}}L\int_{0}^{z(r)}\frac{dz}{\sqrt{r(r-r_{sec})}}\\
& &
\tau(r)=\pm\sqrt{\frac{2}{M}}\int_{0}^{z(r)}\frac{r^{3/2}dz}{\sqrt{r-r_{sec}}}
\end{eqnarray}
if $E=1$.


\pagebreak

\begin{thebibliography}{10}
\bibitem{ON} K.~Oohara and T.~Nakamura, Prog. Theor. Phys. {\bf 71}, 91 (1984).
\bibitem{O} K.~Oohara, Prog. Theor. Phys. {\bf 71}, 738 (1984).
\bibitem{DS} S. L. Detweiler and E. Szedenits, Ap. J. {\bf 231}, 211 (1979).
\bibitem{SP} 
R. F. Stark and T. Piran,  Phys. Rev. Lett. 
	{\bf55},891(1985). Also: Errata,  Phys. Rev. Lett. 
	{\bf56},97(1986).
%
\bibitem{GC} The Grand Challenge Binary Black hole Alliance, see
{\tt http://www.npac.syr.edu/projects/bbh}, also Phys. Rev. Lett.
{\bf 80}, 3915, 2512 and 1812 (1998); Science, {\bf 270}, 941 (1995).
%



\bibitem{KOJI}Y.~Kojima, Prog. Theor. Phys. {\bf 79}, 665 (1988). 
\bibitem{KS1}K.~D.~Kokkotas and B.~Schutz, Mon. Not. R. Astron. Soc. {\bf 255}, 119 (1992).
\bibitem{angmom} There is a single exception to this rule. The angular momentum
of the fluid is an odd-parity dipole perturbation. This perturbation, however, is 
not radiatable, and is irrelevant to the considerations of this paper.
\bibitem{CHANDRAFER}S.~Chandrasekhar and V.~Ferrari, Proc. R. Soc. Lond. A {\bf 434}, 449 (1991).
\bibitem{AKK}N.~Andersson, Y.~Kojima and K.~Kokkotas, Ap. J. {\bf 462}, 855 (1996).
\bibitem{AK2}N.~Andersson and K.~D.~Kokkotas, Mon.~ Not.~ R.~ Astron.~ Soc.
{\bf 297}, 493 (1998).
\bibitem{BBFER}O.~Benhar, E.~Berti and V.~Ferrari, gr-qc/9901037.
\bibitem{KANDER}K.~D.~Kokkotas, T.~A.~Apostolatos and N.~Andersson, gr-qc/9901072.
\bibitem{AK1}N.~Andersson and K.~D.~Kokkotas, Phys. Rev. Lett. {\bf 77}, 4134 (1996).
\bibitem{TOMSAIMAE}K.~Tominaga, M.~Saijo and K.~Maeda,gr-qc/9901040.
\bibitem{FERGUBO}V.~Ferrari, L.~Gualtieri and A.~Borrelli, gr-qc/9901060.
\bibitem{TOOPER}R.~F.~Tooper, Ap. J. {\bf 140}, 434 (1964).
\bibitem{ZERI}F.~J.~Zerilli, Phys. Rev. D {\bf 2}, 2141 (1970). 
\bibitem{THOCAMP}K.~Thorne and A.~Campolattaro, Ap. J. {\bf 149}, 591 (1967).
\bibitem{rw}T.~Regge and J.~A.~Wheeler, Phys.\ Rev. {\bf108}, 1063 (1957).
\bibitem{moncrief}V. Moncrief, Ann. Phys. (NY) {\bf 88},  323  (1974).
\bibitem{CUNPRIMO}C.~T.~Cunningham, R.~H.~Price and V.~Moncrief, Ap. J {\bf 224}, 643 (1978). 
\bibitem{LOUPRICE}C.~O.~Lousto and R.~H.~Price, Phys. Rev. D {\bf 55}, 2124 (1997).
\bibitem{nollertprice} H.-P. Nollert and R. H. Price, J. Math. Phys. {\bf 40}, 980 (1999).
\bibitem{KJELL}K.~Rosquist, Phys. Rev. D {\bf 59}, 044022 (1999).
\bibitem{LIND}L.~Lindblom, Phys. Rev. D {\bf 58}, 024008 (1998).
\bibitem{MTW}C.~W.~Misner, K.~S.~Thorne, and J.~A.~Wheeler, {\it Gravitation} 
(W. H. Freeman, San Francisco, 1973).
\bibitem{CHANDRA}S. Chandrasekhar,{\it The mathematical theory of black holes}
, (Oxford Univ. Press, Oxford, 1983).
\end{thebibliography}
\end{document}